\documentclass[letterpaper, 11pt]{article}
\usepackage{amssymb}
\usepackage{amsthm}
\usepackage{amsmath}
\usepackage{fullpage}
\usepackage{epsfig}
\usepackage{hyperref,txfonts,mathrsfs}

\DeclareSymbolFont{symbolsC}{U}{txsyc}{m}{n}
\SetSymbolFont{symbolsC}{bold}{U}{txsyc}{bx}{n}
\DeclareFontSubstitution{U}{txsyc}{m}{n}
\DeclareMathSymbol{\coloneqq}{\mathrel}{symbolsC}{66}

\newcommand\remove[1]{}

\newcommand{\rnote}[1]{}

\newcommand{\Lip}{\mathrm{Lip}}

\newcommand{\dist}{\mathrm{dist}}

\newcommand{\Z}{\mathbb{Z}}
\newcommand{\R}{\mathbb{R}}
\newcommand{\M}{\mathscr{M}}
\newcommand{\F}{\mathscr{F}}
\newcommand{\ph}{\varphi}
\newcommand{\Id}{\mathrm{Id}}

\newcommand{\Prob}{\mathscr{P}}
\newcommand{\C}{\mathbb{C}}

\newcommand{\e}{\varepsilon}

\newcommand{\polylog}{{\mathrm{polylog}}}

\newcommand{\supp}{{{\mathrm{supp}}}}

\newcommand{\poly}{\mathrm{poly}}

\newtheorem{theorem}{Theorem}[section]
\newtheorem{lemma}[theorem]{Lemma}
\newtheorem{prop}[theorem]{Proposition}

\newtheorem{remark}{Remark}[section]

\begin{document}


\title{Planar Earthmover is not in $L_1$}
\author{
 Assaf Naor\\Microsoft Research\\ {\tt anaor@microsoft.com}\and Gideon Schechtman\thanks{Supported in part by the Israel Science
Foundation.}\\ Weizmann Institute and Microsoft Research\\{\tt
gideon@weizmann.ac.il}}

\date{}
\maketitle

\begin{abstract}
We show that any $L_1$ embedding of the transportation cost (a.k.a.
Earthmover) metric on probability measures supported on the grid
$\{0,1,\ldots,n\}^2\subseteq \R^2$ incurs distortion
$\Omega\left(\sqrt{\log n}\right)$. We also use Fourier analytic
techniques to construct a simple $L_1$ embedding of this space which
has distortion $O(\log n)$.
\end{abstract}

\section{Introduction}
\remove{ Let $d$ be a metric on $\{1,\ldots,n\}$ and
$p=(p_1,\ldots,p_n),q=(q_1,\ldots,q_n)$ be two probability vectors.
The transportation cost distance (also known as the Earthmover
distance in the computer vision/graphics literature) between $p$ and
$q$ is defined as
$$
\tau(p,q)\coloneqq \min\left\{\sum_{i,j=1}^na_{ij}d(i,j):\ \forall\
 i,j\ a_{ij}\ge 0,\ p=(1,\ldots,1)A,\ q=(1,\ldots,1)A^t
\right\}
$$
}

For a finite metric space $(X,d_X)$ we denote by $\mathscr P_X$ the
space of all probability measures on $X$. The transportation cost
distance (also known as the Earthmover distance in the computer
vision/graphics literature) between two probability measures
$\mu,\nu \in \mathscr P_X$ is defined by
$$
\tau(\mu,\nu)=\min\left\{\sum_{x,y\in X} d_X(x,y)\pi(x,y):\ \forall
x,y\in X,\  \pi(x,y)\ge 0,\ \sum_{z\in X} \pi(x,z)=\mu(x), \
\sum_{z\in X}\pi(z,y)=\nu(y)\right\}.
$$
Observe that if $\mu$ and $\nu$ are the uniform probablity
distribution over $k$-point subsets $A\subseteq X$ and $B\subseteq
X$, respectively, then
\begin{eqnarray}\label{eq:def matching}
\tau(\mu, \nu)=\min \left\{\frac{1}{k}\sum_{a\in A} d_X(a,f(a)):\
f:A\to B \ \mathrm{is\  a\  bijection} \right\}.
\end{eqnarray}
This quantity is also known as the {\em minimum weight matching}
between $A$ and $B$, corresponding to the weight function
$d_X(\cdot,\cdot)$ (see~\cite{PWR89}). Thus, the Earthmover distance
is a natural measure of similarity between
images~\cite{PWR89,GRT98,GRT00}- the distance is  the optimal way to
match various features, where the cost of such a matching
corresponds to the sum of the distances between the features that
were matched. Indeed, such metrics occur in various contexts in
computer science: Apart from being a popular distance measure in
graphics and vision~\cite{PWR89,GRT98,GRT00,IT03},  they are used as
LP relaxations for classification problems such as $0$-extension and
metric labelling~\cite{CKNZ01,Cha02,AFHKTT04}. Transportation cost
metrics are also prevalent in several areas of analysis and PDE (see
the book~\cite{Villani03} and the references therein).

Following extensive work on nearest neighbor search and data stream
computations for $L_1$ metrics
(see~\cite{InM98,Ind00,Ind01,DIIM04,Ind04-handbook}), it became of
great interest to obtain low distortion embeddings of useful metrics
into $L_1$ (here, and in what follows, $L_1$ denotes the space of
all Lebesgue measurable functions $f:[0,1]\to \R$, such that
$\|f\|_1\coloneqq \int_0^1|f(t)|dt<\infty$). Indeed, such embeddings
can be used to construct approximate nearest neighbor databases,
with an approximation guarantee depending on the {\em distortion} of
the embedding (we are emphasizing here only one aspect of the
algorithmic applications of low distortion embeddings into $L_1$-
they are also crucial for the study of various cut problems in
graphs, and we refer the reader
to~\cite{Mat01,IndMat04-handbook,Ind04-stoc} for a discussion of
these issues).

In the context of the Earthmover distance, nearest neighbor search
(a.k.a. similarity search in the vision literature) is of
particular importance. It was therefore asked (see,
e.g.~\cite{jiriproblems}) whether the Earthmover distance embeds
into $L_1$ with constant distortion (the best known upper bounds
on the $L_1$ distortion were obtained in~\cite{Cha02,IT03}, and
will be discussed further below). In~\cite{KN05} the case of the
Hamming cube was settled negatively: It is shown there that any
embedding of the Earthmover distance on $\{0,1\}^d$ (equipped with
the $L_1$ metric) incurs distortion $\Omega(d)$. However, the most
interesting case is that of the Earthmover distance on $\R^2$, as
this corresponds to a natural similarity measure between
images~\cite{GRT00} (indeed, the case of the $L_1$ embeddability
of planar Earthmover distance was explicitly asked
in~\cite{jiriproblems}). Here we settle this problem negatively by
obtaining the first super-constant lower bound on the $L_1$
distortion of the planar Earthmover distance. To state it we first
recall some definitions.

Given two metric spaces $(X,d_X)$ and $(Y,d_Y)$, and a mapping
$f:X\to Y$, we denote its Lipschitz constant by
$$
\|f\|_{\Lip}\coloneqq \sup_{\substack{x,y\in X\\x\neq y}}
\frac{d_Y(f(x),f(y))}{d_X(x,y)}.
$$
If $f$ is one to one then its distortion is defined as
$$
\dist(f)\coloneqq \|f\|_{\Lip}\cdot
\|f^{-1}\|_{\Lip}=\sup_{\substack{x,y\in X\\x\neq y}}
\frac{d_Y(f(x),f(y))}{d_X(x,y)}\cdot\sup_{\substack{x,y\in
X\\x\neq y}} \frac{d_X(x,y)}{d_Y(f(x),f(y))}.
$$
The smallest distortion with which $X$ can be embedded into $Y$ is
denoted $c_Y(X)$, i.e.,
$$
c_Y(X)\coloneqq \inf\left\{\dist(f):\ f:X\hookrightarrow Y\
\mathrm{is\ one\ to\ one} \right\}.
$$
When $Y=L_p$ we use the shorter notation $c_Y(X)=c_p(X)$. Thus,
the parameter $c_2(X)$ is the Euclidean distortion of $X$ and
$c_1(X)$ is the $L_1$ distortion of $X$.

Our main result bounds from below the $L_1$ distortion of the
space of probability measures on the $n$ by $n$ grid, equipped
with the transportation cost distance.

\begin{theorem}\label{thm:main}
$c_1\left(\Prob_{\{0,1,\ldots,n\}^2},\tau\right)=\Omega\left(\sqrt{\log
n} \right)$.
\end{theorem}

After reducing the problem to a functional analytic question, our
proof of Theorem~\ref{thm:main} is a discretization of a theorem
of Kislyakov from 1975~\cite{Kis75}. We attempted to make the
presentation self contained by presenting here appropriate
versions of the various functional anlaytic lemmas that are used
in the proof.

For readers who are more interested in the minimum cost matching
metric~\eqref{eq:def matching}, we also prove the following lower
bound:

\begin{theorem}[Discretization]\label{thm:discretization}
For arbitrarily large integers $n$ there is a family $\mathscr Y$ of
disjoint $n$-point subsets of $\left\{0,1\ldots,n^3\right\}^2$, with
$|\mathscr Y|\le n^{O(\log \log n)}$, such that any $L_1$ embedding
of $\mathscr Y$, equipped with the minimum weight matching metric
$\tau$, incurs distortion
$$
\Omega\left(\sqrt{\log \log \log n}\right)=\Omega\left(\sqrt{\log
\log \log |\mathscr Y| }\right)
$$
\end{theorem}

A metric spaces $(X,d_X)$ is said to embed into squared $L_2$, or to
be of negative type, if the metric space $\left(X,\sqrt{d_X}\right)$
is isometric to a subset of $L_2$. Squared $L_2$ metrics are
important in various algorithmic applications since it is possible
to efficiently solve certain optimization problems on them using
semidefinite programming (see the discussion in~\cite{ALN05,KV05}).
It turns out that planar Earthmover does not embed into any squared
$L_2$ metric:

\begin{theorem}[Nonembeddability into squared
$L_2$]\label{thm:negative type} $ \lim_{n\to\infty}
c_2\left(\Prob_{\{0,\ldots,n\}^2},\sqrt{\tau}\right)=\infty. $
\end{theorem}

\bigskip

Motivated by the proof of Theorem~\ref{thm:main}, we also construct
simple low-distortion embeddings of the space
$\left(\Prob_{\{0,1,\ldots,n\}^2},\tau\right)$ into $L_1$. It is
convenient to work with probability measures on the torus $\Z_n^2$
instead of the grid $\{0,1,\ldots,n\}^2$. One easily checks that
$\{0,\ldots,n\}^2$ embeds with constant distortion into $\Z_{2n}^2$
(see e.g. Lemma 6.12 in~\cite{MN05}). Every $\mu\in \Prob_{\Z_n^2}$
can be written in the Fourier basis as
\begin{eqnarray}\label{eq:decompose}
\mu=\sum_{(u,v)\in \Z_n^2} \widehat \mu(u,v) e_{uv},
\end{eqnarray}
where $$ \forall (a,b),(u,v)\in \Z_n^2,\  e_{uv}(a,b)\coloneqq
e^{\frac{2\pi i(au+bv)}{n}},\quad \mathrm{and}\quad \forall (u,v)\in
\Z_n^2,\ \  \widehat \mu(u,v)\coloneqq\frac{1}{n^2} \sum_{(a,b)\in
\Z_n^2} \mu(a,b)e_{uv}(-a,-b).
$$
Observe that for $n=2^k+1$, $k\in \mathbb N$, the
decomposition~\eqref{eq:decompose} can be computed in time
$O\left(n^2\log n\right)$ using the fast Fourier
transform~\cite{Rock94}. Motivated in part by the results
of~\cite{Pel89} (see also~\cite{BBPW01,PW03}), we define
\begin{eqnarray}\label{eq:def A}
 A\mu=\sum_{(u,v)\in \Z_n^2\setminus \{(0,0)\}} \frac{e^{\frac{2\pi
i u}{n}}-1}{\big|e^{\frac{2\pi i u}{n}}-1\big|^2+\big|e^{\frac{2\pi
i v}{n}}-1\big|^2}\cdot \widehat \mu(u,v)\cdot e_{uv},
\end{eqnarray}
and
\begin{eqnarray}\label{eq:def B}
B\mu=\sum_{(u,v)\in \Z_n^2\setminus \{(0,0)\}} \frac{e^{\frac{2\pi i
v}{n}}-1}{\big|e^{\frac{2\pi i u}{n}}-1\big|^2+\big|e^{\frac{2\pi i
v}{n}}-1\big|^2}\cdot \widehat \mu(u,v)\cdot e_{uv}.
\end{eqnarray}

\begin{theorem}\label{thm:upper} The mapping
$\mu\mapsto (A\mu,B\mu)$ from $\left(\Prob_{\Z_n^2},\tau\right)$
to $L_1\left(\Z_n^2\right)\oplus L_1\left(\Z_n^2\right)$ is
bi-Lipschitz, with distortion $O(\log n)$.
\end{theorem}
The $O(\log n)$ distortion  in Theorem~\ref{thm:upper} matches the
best known distortion guarantee proved in~\cite{IT03,Cha02}. But,
our embedding has various new features. First of all, it is a {\em
linear} mapping into a low dimensional $L_1$ space, which is based
on the computation of the Fourier transform. It is thus very fast
to compute, and is versatile in the sense that it might behave
better on images whose Fourier transform is sparse (we do not
study this issue here). Thus there is scope to apply the embedding
on certain subsets of the frequencies, and this might improve the
performance in practice. This is an interesting ``applied"
question which should be investigated further (see the
``Discussion and open problems" section).

\section{Preliminaries and notation}

For the necessary background on measure theory we refer to the
book~\cite{Rudin87}, however, in the setting of the present paper,
our main results will deal with finitely supported measures, in
which case no background and measurabilty assumptions are
necessary. We also refer to the book~\cite{Villani03} for
background on the theory of optimal transportation of measures.
Let $(X,d_X)$ be a metric space. We denote by $\M_X$ the space of
all Borel measures on $X$ with bounded total variation, and by
$\Prob_X\subseteq \M_X$ the set of all Borel {\em probability}
measures on $X$. We also let $\M^+_X\subseteq \M_X$ be the space
of {\em non-negative} measures on $X$ with finite total mass, and
we denote by $\M^0_X\subseteq \M_X$
 the space of all measures $\mu\in \M_X$ with $\mu(X)=0$. Given a
measure $\mu\in \M_X$, we can decompose it in a unique way as
$\mu=\mu^+-\mu^-$, where $\mu^+,\mu^-\in \M_X^+$ are disjointly
supported. If $\mu,\nu\in \M_X^+$ have the same total mass, i.e.
$\mu(X)=\nu(X)<\infty$, then we let $\Pi(\mu,\nu)$ be the space of
all {\em couplings} of $\mu$ and $\nu$, i.e. all non-negative Borel
measures $\pi$ on $X\times X$ such that for every measurable bounded
$f:X\to \R$,
$$
\int_{X\times X} f(x)d\pi(x,y)=\int_X
f(x)d\mu(x),\quad\mathrm{and}\quad \int_{X\times X}
f(y)d\pi(x,y)=\int_X f(y)d\nu(y).
$$
Observe that in the case of finitely supported measures, this
condition translates to the standard formulation, in which we
require that the marginals of $\pi$ are $\mu$ and $\nu$, i.e.
$$
\forall x,y\in X,\ \sum_{z\in X} \pi(x,z)=\mu(x),\quad
\mathrm{and}\quad \sum_{z\in X}\pi(z,y)=\nu(y).
$$
The {\em transportation cost distance} between $\mu$ and $\nu$,
denoted here by $\tau(\mu,\nu)=\tau_{(X,d_X)}(\mu.\nu)$ (and also
referred to in the literature as the Wasserstein $1$ distance,
Monge-Kantorovich distance, or the Earthmover distance), is
\begin{eqnarray}\label{eq:def tau}
\tau(\mu,\nu)\coloneqq \inf\left\{\int_{X\times X} d_X(x,y)\,
d\pi(x,y):\ \pi\in \Pi(\mu,\nu)\right\}.
\end{eqnarray}
\remove{ Observe that if $\mu$ and $\nu$ are the uniform
probablity distribution over $k$-point subsets $A\subseteq X$ and
$B\subseteq X$, respectively, then
$$
\tau(\mu, \nu)=\min \left\{\frac{1}{k}\sum_{a\in A} d_X(a,f(a)):\
f:A\to B \ \mathrm{is\  a\  bijection} \right\}.
$$
(This quantity is also known as the {\em minimum weight matching}
between between $A$ and $B$, corresponding to the weight function
$d_X(\cdot,\cdot)$.) }
 For $\mu\in \M_X^0$, $\mu^+(X)=\mu^-(X)$,
so we may write $\|\mu\|_{\tau}\coloneqq \tau(\mu^+,\mu^-)$. This
is easily seen to be a norm on the vector space
$\M_{X,\tau}^0\coloneqq \left\{\mu\in \M_X^0:\
\|\mu\|_{\tau}<\infty\right\}$.

Fix some $x_0\in X$, and let $\Lip_0(X)=\Lip_{x_0}(X)$ be the linear
space of all Lipschitz mappings $f:X\to \R$ with $f(x_0)=0$,
equipped with the norm $\|\cdot \|_{\Lip}$ (i.e. the norm of a
function equals its Lipschitz constant). Any $\mu\in \M_{X,\tau}^0$
can be thought of as a bounded linear functional on $\Lip_0(X)$,
given by $f\mapsto \int_{X}f d\mu$. The famous {\em Kantorovich
duality theorem} (see Theorem 1.14 in~\cite{Villani03}) implies that
$\Lip_0(X)^*=\M_{X,\tau}^0$, in the sense that every bounded linear
functional on $\Lip_0(X)$ is obtained in this way, and for every
$\mu\in \M_{X,\tau}^0$,
$$
\|\mu\|_{\tau}=\|\mu\|_{\Lip_0(X)^*}\coloneqq
\sup\left\{\int_{X}fd\mu:\  f\in \Lip_0(X),\ \|f\|_{\Lip}\le
1\right\}.
$$
(We note that this identity amounts to duality of linear
programming.)

\section{Proof of Theorem~\ref{thm:main}}

Fix an integer $n\ge 2$ and denote $X=\{0,1,\ldots,n-1\}^2$,
equipped with the standard Euclidean metric. In what follows, for
concreteness,  $\Lip_0\coloneqq \Lip_0(X)$ is defined using the base
point $x_0=(0,0)$. Also, for ease of notation we denote
$\M=\M_{X,\tau}^0$. Observe that $Lip_0$ and $\M$ are vector spaces
of dimension $n^2-1$, and by Kantorovich duality, $\Lip_0^*=\M$ and
$\M^*=\Lip_0$.

Assume that $F:\Prob_X\to L_1$ is a bi-Lipschitz embedding,
satisfying for all two probability measures $\mu,\nu\in \Prob_X$,
\begin{equation}\label{eq:L}
\tau(\mu,\nu)\le \|F(\mu)-F(\nu)\|_1\le L\cdot \tau(\mu,\nu).
\end{equation}
Our goal is to bound $L$ from below. We begin by reducing the
problem to the case of {\em linear mappings}. Recall that given two
normed spaces $(Z,\|\cdot\|_Z)$ and $(W,\|\cdot\|_W)$, the norm of a
linear mapping $T:Z\to W$ is defined as $\|T\|=\sup_{z\in
Z\setminus\{0\}}\frac{\|Tz\|_W}{\|z\|_Z}$ (observe that in this case
$\|T\|=\|T\|_{\Lip}$).

\begin{lemma}[Reduction to a linear embedding of $\M$ into $\ell_1^N$]\label{lem:M in
Prob} Under the assumption of an existence of an embedding
$F:\Prob_X\to L_1$ satisfying~\eqref{eq:L}, there exists an integer
$N$, and an invertible linear operator $T: \M\to \ell_1^N$, with
$\|T\|\le 2L$ and $\|T(\mu)\|_1\ge \|\mu\|_\tau$ for all $\mu\in \M$
(the factor $2$ can be replaced by $1+\e$ for every $\e>0$, but this
is irrelevant for us here).
\end{lemma}

\begin{proof} By translation we may assume that $F$ maps the uniform
measure on $X$ to $0$. For $\mu\in \M$ denote
$\|\mu\|_\infty\coloneqq \max_{x\in X} |\mu(x)|$. Observe that it is
always the case that $\|\mu\|_\infty\le \|\mu\|_\tau$. Indeed, if
$\pi\in \Pi(\mu^+,\mu^-)$ then
$$
\int_{X\times X} \|x-y\|_2d\pi(x,y)\ge \int_{X\times X}
d\pi(x,y)=\mu^+(X)=\mu^-(X)\ge \|\mu\|_\infty.
$$
Let $B_\M$ denote the unit ball of $\M$. Define for $\mu \in B_\M$ a
probability measure $\psi(\mu)\in \Prob(X)$ by
$\psi(\mu)(x)\coloneqq\frac{\mu(x)+1}{n^2}$. It is clear that for
every $\mu,\nu\in \M$,
$\|\mu-\nu\|_\tau=\frac{1}{n^2}\cdot\|\psi(\mu)-\psi(\nu)\|_{\tau}$.
The mapping $h\coloneqq n^2\cdot F\circ \psi: B_\M\to L_1$ satisfies
$h(0)=0$, $\|h\|_{\Lip}\le L$, and $\|h(\mu)-h(\nu)\|_1\ge
\|\mu-\nu\|_\tau$. This implies that there exists a map $\tilde
h:\M\to L_1$ satisfying the same inequalities. We shall present two
arguments establishing this fact: The first is a soft
non-constructive proof, using the notion of ultraproducts, and the
second argument is more elementary, but does not preserve the
Lipschitz constant.

Let $\mathscr U$ be a free ultrafilter on $\mathbb N$, and denote by
$(L_1)_{\mathscr U}$ the corresponding ultrapower of $L_1$
(see~\cite{Hein80} for the necessary background on ultrapowers of
Banach spaces. In particular, it is shown there that
$(L_1)_{\mathscr U}$ is isometric to an $L_1(\sigma)$ space, for
some measure $\sigma$). Define for $\mu\in \M$, $\tilde
h(\mu)=\left(j\cdot h(\mu/j)\right)_{j=1}^\infty/\mathscr U$, where
we set, say, $h(\nu)=0$ for $\nu\in \M\setminus B_\M$. Then, by
standard arguments, $\|\tilde h\|_{\Lip}\le L$ and $\|\tilde
h^{-1}\|_{\Lip}\le 1$. Moreover, $\tilde h (\M)$ spans a separable
subspace of $(L_1)_{\mathscr U}$, and thus we may assume without
loss of generality that $\tilde h$ takes values in $L_1$.

An alternative proof (for those of us who don't mind losing a
constant factor), proceeds as follows. For every $f\in L_1$ let
$\chi(f): [0,1]\times \R\to \{-1,0,1\}$ be the function given by
$$
\chi(f)(s,t)=\mathrm{sign}(f(s))\cdot {\bf
1}_{\big[0,|f(s)|\big]}(t)=\left\{\begin{array}{ll} 1 & f(s)>0,\ 0\le t\le f(s),\\
-1 & f(s)<0,\ 0\le t\le -f(s),\\ 0 & \mathrm{otherwise}.
\end{array}\right.
$$
It is straightforward to check that
$\|\chi(f)-\chi(g)\|_{L_1([0,1]\times \R)}= \|f-g\|_1$ for every
$f,g\in L_1$ (We note here that the space $L_1([0,1]\times \R)$ is
isometric to $L_1$.)
\remove{ Define $\tilde h: \M\to
L_1([0,1]\times \R)$ by
$$
\tilde h(\mu)=\left\{\begin{array}{ll} \chi\circ h(\mu),& \mu\in B_\M\\
 \|\mu\|_\tau\cdot  \chi\circ h(\mu/\|\mu\|_\tau) & \mu\in \M\setminus B_\M.
\end{array}\right.
$$
}
 Define $\tilde h: \M\to
L_1([0,1]\times \R)$ by setting $\tilde h (\mu)=\|\mu\|_\tau\cdot
\chi\circ h(\mu/\|\mu\|_\tau)$ for $\mu\in \M\setminus\{0\}$, and
$\tilde h(0)=0$. Since for every $f\in L_1$, $\chi(f)$ takes values
in $\{-1,0,1\}$, we have the following pointwise identity for every
$\mu,\nu\in \M$ with $\|\mu||_\tau\ge \|\nu\|_\tau$:
$$
\left|\tilde h(\mu)-\tilde h(\nu)\right|= \|\nu\|_\tau\cdot
\left|\chi\circ h\left(\frac{\mu}{\|\mu\|_\tau}\right)- \chi\circ
h\left(\frac{\nu}{\|\nu\|_\tau}\right)\right|+\left(\|\mu\|_\tau-\|\nu\|_\tau\right)\cdot
\left|\chi\circ h\left(\frac{\mu}{\|\mu\|_\tau}\right)\right|.
$$
Thus
\begin{eqnarray}\label{eq:L_1 identity}
\left\|\tilde h(\mu)-\tilde h(\nu)\right\|_{L_1([0,1]\times
\R)}&=&\|\nu\|_\tau\cdot \left\|
h\left(\frac{\mu}{\|\mu\|_\tau}\right)-
h\left(\frac{\nu}{\|\nu\|_\tau}\right)\right\|_1+\left(\|\mu\|_\tau-\|\nu\|_\tau\right)\cdot
\left\| h\left(\frac{\mu}{\|\mu\|_\tau}\right)\right\|_1\\
&\ge& \nonumber  \|\nu\|_\tau\cdot \left\| \frac{\mu}{\|\mu\|_\tau}-
\frac{\nu}{\|\nu\|_\tau}\right\|_\tau+\|\mu\|_\tau-\|\nu\|_\tau\\
&\ge& \nonumber
\|\nu-\mu\|_\tau-\left\|\mu-\frac{\|\nu\|_\tau}{\|\mu\|_\tau}\mu\right\|_\tau+\|\mu\|_\tau-\|\nu\|_\tau\\
&=&\nonumber \|\nu-\mu\|_{\tau}.
\end{eqnarray}
It also follows from the identity~\eqref{eq:L_1 identity} that
\begin{eqnarray*}
\left\|\tilde h(\mu)-\tilde h(\nu)\right\|_{L_1([0,1]\times
\R)}&\le& L\|\nu\|_\tau\cdot
\left\|\frac{\mu}{\|\mu\|_\tau}-\frac{\nu}{\|\nu\|_\tau}\right\|_\tau+L\|\mu-\nu\|_\tau\\
&\le&L\|\mu-\nu\|_\tau+L\|\nu\|_{\tau}\|\mu\|_\tau\cdot
\left|\frac{1}{\|\mu\|_\tau}-\frac{1}{\|\nu\|_\tau}\right|+L\|\mu-\nu\|_\tau\\
&\le& 3L\|\mu-\nu\|_\tau.
\end{eqnarray*}

We are now in position to use a Theorem of Ribe~\cite{Ribe76} (see
also~\cite{HM82}, and Corollary 7.10 in~\cite{BL00}, for softer
proofs), which implies that there is an into linear isomorphism
$S:\M\to L_1^{**}$ satisfying $\|S\|\le L$ and $\|S^{-1}\|\le 1$.
Since $\M$ is finite dimensional, by the principle of local
reflexivity~\cite{LR69} (alternatively by Kakutani's
representation theorem~\cite{Kak41,LTII77}), and a simple
approximation argument, we get that there exists an integer $N$
and an into linear isomorphism $T:\M\to \ell_1^N$ satisfying
$\|T\|\le 2L$ and $\|T^{-1}\|\le 1$ (the value of $N$ is
irrelevant for us here, and indeed it is possible to conclude the
proof without passing to a finite dimensional $L_1$ space, but
this slightly simplifies some of the ensuing arguments. For
completeness we note here that using a theorem of
Talagrand~\cite{Tal90} we can ensure that $N=O(n\log n)$).
\end{proof}

\bigskip

From now on let $T:\M\to \ell_1^N$ be the linear operator guaranteed
by Lemma~\ref{lem:M in Prob}. Since $T$ is an isomorphism, the
adjoint operator $T^*:\ell_\infty^N\to \M^*=\Lip_0$ is a quotient
mapping, i.e. $\|T^*\|\le 2L$ and the image of the unit ball of
$\ell_\infty^N$ under $T^*$ contains the unit ball of $\Lip_0$. We
now define three more auxiliary linear operators. The first is the
formal identity $\Id:\Lip_0\to W$, where $W$ is the space of all
functions $f:X\to \R$ with $f(0)=0$, equipped with the (discrete
Sobolev) norm
$$
\|f\|_W\coloneqq  \sum_{i=0}^{n-1}\sum_{j=0}^{n-2}
|f(i,j+1)-f(i,j)|+  \sum_{j=0}^{n-1}\sum_{i=0}^{n-2}
|f(i+1,j)-f(i,j)|.
$$
The second operator is also a formal identity (discrete Sobolev
embedding) $S:W\to \ell_2(X)$, where the Euclidean norm on
$\ell_2(X)$ is taken with respect to the counting measure on $X$.
The final operator we will use is the Fourier operator $\F:
\ell_2(X)\to \ell_2(X)$, defined for $f:X\to \R$ by
$$
\F(f)(u,v)\coloneqq \frac{1}{n^2}\sum_{(k,\ell)\in X}
f(k,\ell)\sin\left(\frac{2\pi uk}{n}\right)\cdot
\sin\left(\frac{2\pi v\ell}{n}\right).
$$
The following lemma summarizes known estimates on the norms of these
operators:

\begin{lemma}[Operator norm bounds]\label{lem:opnorm} The following
operator norm bounds hold true: \\ \textbullet\  $\|\Id\|\le
2n(n-1)$.\quad \textbullet\ $\|S\|\le 1$.\quad\textbullet\
$\|\F\|\le \frac{1}{n}$.

\end{lemma}

\begin{proof} The first statement means that for every $f:X\to \R$ with $f(0)=0$,
$\|f\|_W\le 2n(n-1)\|f\|_{\Lip}$, which is obvious from the
definitions. The second assertion is that $\|f\|_{2}\le \|f\|_W$.
This is a discrete version of Sobolev's inequality~\cite{PW03}
(with non-optimal constant), which can be proved as follows. First
of all, since $f(0)=0$, for every $(u,v)\in X$,
\begin{eqnarray}\label{eq:first integration}
|f(u,v)|&=& \left|\sum_{k=0}^{u-1}\nonumber
\left[f(k+1,v)-f(k,v)\right]+
\sum_{\ell=0}^{v-1}\left[f(0,\ell+1)-f(0,\ell)\right]\right|\\&\le&
\sum_{k=0}^{n-2} \left|f(k+1,v)-f(k,v)\right|+
\sum_{\ell=0}^{n-2}\left|f(0,\ell+1)-f(0,\ell)\right|\coloneqq A(v).
\end{eqnarray}
Analogously,
\begin{eqnarray}\label{eq:second integration}
|f(u,v)|\le \sum_{\ell=0}^{n-2} \left|f(u,\ell+1)-f(u,\ell)\right|+
\sum_{k=0}^{n-2}\left|f(k+1,0)-f(k,0)\right|\coloneqq B(u).
\end{eqnarray}
Multiplying~\eqref{eq:first integration} and~\eqref{eq:second
integration}, and summing over $X$, we see that
\begin{eqnarray*}
\|f\|_{2}^2\le \sum_{(u,v)\in X} A(v)B(u)= \Biggl(\sum_{v=0}^{n-1}
A(v)\Biggr)\cdot \Biggl(\sum_{u=0}^{n-1} B(u)\Biggr) \le
\frac14\Biggl(\sum_{v=0}^{n-1} A(v)+\sum_{u=0}^{n-1} B(u)\Biggr)^2
\le \frac14 \left(2\|f\|_W\right)^2.
\end{eqnarray*}

The final assertion follows from the fact that the system of
functions $\left\{(k,\ell)\mapsto \sin\left(\frac{2\pi
uk}{n}\right)\cdot \sin\left(\frac{2\pi
v\ell}{n}\right)\right\}_{(u,v)\in X}$ are orthogonal in $\ell_2(X)$
and have norms bounded by $n$.
\end{proof}

We now recall some facts related to absolutely summing operators on
Banach spaces (we refer the interested reader to~\cite{Tom89,Woj96}
for more information on this topic). Given two Banach spaces $Y$ and
$Z$, the $\pi_1$ norm of an operator $A:Y\to Z$, denoted $\pi_1(A)$,
is defined to be the smallest constant $K>0$ such that for every
$m\in \mathbb N$ and every $y_1,\ldots,y_m\in Y$ there exists a norm
$1$ linear functional $y^*\in Y^*$ satisfying
\begin{eqnarray}\label{eq:def pi1}
\sum_{j=1}^m \|Ay_j\|_Z\le K \sum_{j=1}^m |y^*(y_j)|.
\end{eqnarray}
This defines an {\em ideal norm} in the sense that it is a norm, and
for every two operators $P:W\to Y$ and $Q:Z\to V$ we have
$\pi_1(QAP)\le \|Q\|\cdot\pi_1(A)\cdot \|P\|$. Observe that it is
always the case that $\pi_1(A)\ge \|A\|$.

\begin{lemma}\label{lem:pi1} Using the above notation,
$\pi_1(\Id)\le 2n(n-1) $. Therefore, Lemma~\ref{lem:opnorm} implies
that $$\pi_1(\F\circ S\circ \Id\circ T^*)\le 4nL.$$
\end{lemma}

\begin{proof} Fix $f_1,\ldots,f_m:X\to \R$ with $f_1(0)=\cdots
=f_m(0)=0$. Then
\begin{eqnarray*}
\sum_{j=1}^m \|f_j\|_W&=&\sum_{s=0}^{n-1}\sum_{t=0}^{n-2}
\sum_{j=1}^m\big( |f_j(s,t+1)-f_j(s,t)|+|f_j(t+1,s)-f_j(t,s)|\big)\\
&\le& 2n(n-1)\max\left\{\max_{\substack{0\le s\le n-1\\0\le t\le
n-2}}\sum_{j=1}^m |f_j(s,t+1)-f_j(s,t)|, \max_{\substack{0\le s\le
n-1\\0\le t\le n-2}}\sum_{j=1}^m|f_j(t+1,s)-f_j(t,s)|\right\}.
\end{eqnarray*}
Assume without loss of generality that the maximum above equals $
\sum_{j=1}^m |f_j(s_0,t_0+1)-f_j(s_0,t_0)|$, for some $0\le s_0\le
n-1$ and $0\le t_0\le n-2$. Consider the measure
$\mu=\delta_{(s_0,t_0+1)}-\delta_{(s_0,t_0)}\in \M=\Lip_0^*$. One
checks that $\|\mu\|_\tau= 1$, and $\sum_{j=1}^m
|f_j(s_0,t_0+1)-f_j(s_0,t_0)|=\sum_{j=1}^m |\mu(f_j)|$, implying the
required result.
\end{proof}

\bigskip

The fundamental property of the $\pi_1$ norm is the Pietsch
Factorization Theorem (see~\cite{Tom89}), a special case of which is
the following lemma. We present a proof for the sake of
completeness.

\begin{lemma}[Pietsch factorization]\label{lem:pietsch} Let $Y$ be a
Banach space, and fix a linear operator $A:\ell_\infty^N\to Y$. Then
there exists a probability measure $\sigma$ on $\{1,\ldots,N\}$ and
a linear operator $R:L_1(\sigma)\to Y$ such that $A=R\circ I$, where
$I$ is the formal identity from $\ell_\infty^N$ to $L_1(\sigma)$,
and $\|R\|=\pi_1(A)$.
\end{lemma}

\begin{proof}
Recall that $A:\ell_\infty^N\to Y$ satisfies for all
$x_1,\ldots,x_m\in \ell_\infty^m$,
\[
\sum_{i=1}^m\|Ax_i\|\le \pi_1(A)\cdot\sup_{\substack{x^*\in
\left(\ell_\infty^N\right)^*\\ \|x^*\|=1}}\sum_{i=1}^m |x^*(x_i)|=
\pi_1(A)\cdot\max_{1\le k\le N}\sum_{i=1}^m |x_i(k)|,
\]
where the last equality follows from the fact that the evaluation
functionals $x\mapsto x(k)$ are the extreme points of the unit
ball of $\ell_1^N=\left(\ell_\infty^N\right)^*$. Consider the two
subsets of $\R^N$:
\[
K_1=\left\{\Biggl(\sum_{i=1}^m\|Ax_i\|-\pi_1(A)\sum_{i=1}^m
|x_i(k)|\Biggr)_{k=1}^N:\ m\in \mathbb N\ \mbox{and} \
x_1,\ldots,x_m\in\ell_\infty^N\right\},
\]
and
\[
K_2=\left\{x\in\ell_\infty^N:\ x(k)>0 \ \mbox {for all} \ 1\le k \le
N\right\}.
\]
Note that $K_1$ and $K_2$ are disjoint convex cones with $K_2$ open.
It follows from the separation theorem that there is a non zero
$\sigma\in\ell_1^N$ such that $\sigma(x)\le 0$ for all $x\in K_1$
and $\sigma(x)\ge 0$ for all $x\in K_2$. The second inequality
implies that $\sigma$ is positive; we can then assume, by
renormalizing, that it is a probability measure on $\{1,\dots, N\}$.
The first inequality implies that
\[
\|Ax\|\le\pi_1(A)\int_{\{1,\ldots,N\}}|x(k)|d\sigma
\]
for all $x \in \R^N$. Define $Rx=Ax$.
\end{proof}

From now on let $R$ and $\sigma$ be the operator and probability
measure corresponding to $A=\F\circ S\circ \Id\circ T^*$ in
Lemma~\ref{lem:pietsch}. Thus $R\circ I=\F\circ S\circ \Id\circ T^*$
and $\|R\|\le 4nL$. Schematically, we have the following commuting
diagram:

\begin{figure}[h]
\begin{minipage}{3in}
\begin{center}
        \centerline{\quad\quad\quad\quad\quad\quad\quad\quad\quad\quad\quad\quad\quad\quad\quad\quad\quad\quad\quad\quad\quad\quad\quad\hbox{
        \psfig{figure=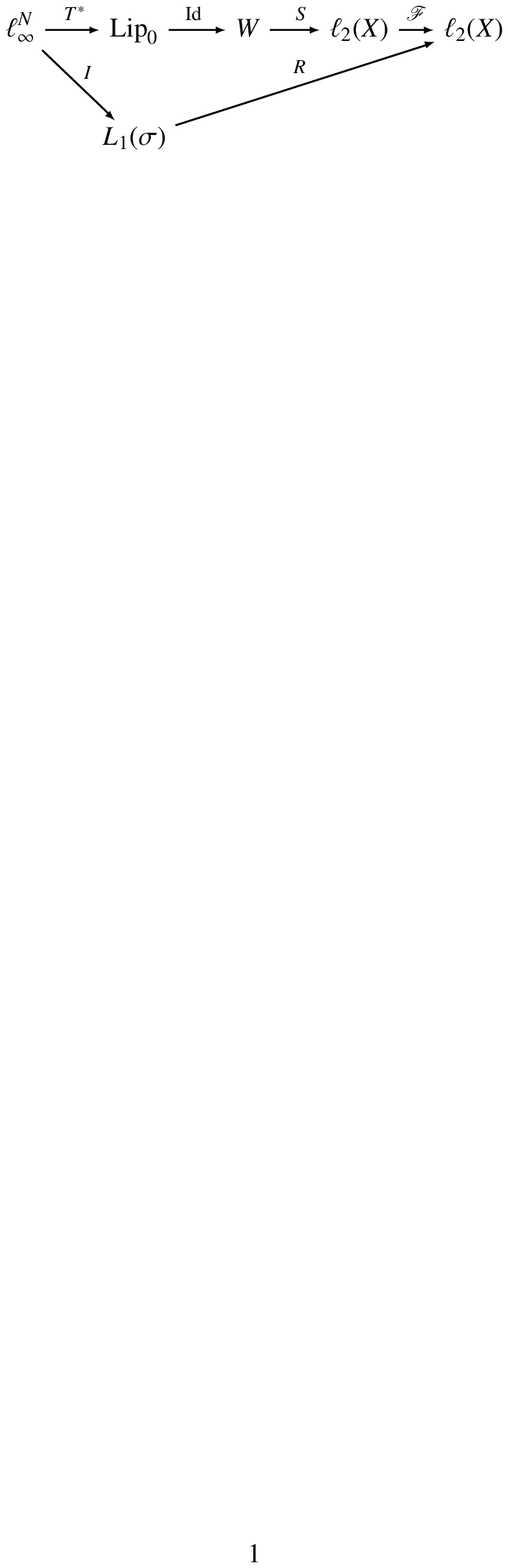,height=1in,width=3in}
        }}
\end{center}
\end{minipage}
\end{figure}

We need only one more simple result from classical Banach space
theory. This is a special case of a more general theorem, but we
shall prove here only what is needed to conclude the proof of
Theorem~\ref{thm:main}.

\begin{lemma}\label{lem:lattice} Let $R:L_1(\sigma)\to \ell_2$ be a linear operator.
Fix $f: \R^N\to [0,\infty)$. Then there is $x\in \ell_2$ with
non-negative coordinates such that
$$
R\left(\left\{g:\R^N\to \R:\ \forall j,\ |g(j)|\le
f(j)\right\}\right)\subseteq \{y\in \ell_2:\ \forall j,\ |y_j|\le
x_j\},
$$
and $\|x\|_2\le \|R\|\cdot \|f\|_{L_1(\sigma)}$.
\end{lemma}

\begin{proof} $R$ is given by a matrix $(R_{ij}:\ i=1,\ldots, N,\
j\in \mathbb N)$. In other words, for every $j$,
$(Rf)_j=\sum_{i=1}^N R_{ij}f(i)$. Observe that using this notation,
\begin{eqnarray}\label{eq:norm}
\|R\|=\max_{1\le i\le N}\,
\Biggl(\frac{1}{\sigma(i)^2}\sum_{j=1}^\infty R_{ij}^2\Biggr)^{1/2}.
\end{eqnarray}

Fix $g\in L_1(\sigma)$ such that for all $i\in \{1,\ldots, N\}$,
$|g(i)|\le f(i)$. Then for all $j$,
$$
|(Rg)_j|\le \sum_{i=1}^N |R_{ij}| f(i)\coloneqq x_j .
$$
Now,
\begin{eqnarray*}
\|x\|_2=\left[\sum_{j=1}^\infty \Biggl(\sum_{i=1}^N |R_{ij}|
f(i)\Biggr)^2\right]^{1/2}\le \sum_{i=1}^N\Biggl(\sum_{j=1}^\infty
|R_{ij}|^2f(i)^2\Biggr)^{1/2} = \sum_{i=1}^n
\sigma(i)f(i)\Biggl(\frac{1}{\sigma(i)^2}\sum_{j=1}^\infty
R_{ij}^2\Biggr)^{1/2}\le \|R\|\cdot \|f\|_{L_1(\sigma)},
\end{eqnarray*}
where we have used~\eqref{eq:norm}.
\end{proof}

\bigskip

We are now in position to conclude the proof of
Theorem~\ref{thm:main}.

\begin{proof}[Proof of Theorem~\ref{thm:main}] For $(u,v)\in \{1,\ldots,n\}^2$
define $\ph_{u,v}: X\to \R$ by
$$
\ph_{u,v}(k,\ell)\coloneqq \frac{1}{u+v}\cdot \sin\left(\frac{2\pi
uk}{n}\right)\cdot \sin\left(\frac{2\pi v\ell}{n}\right).
$$
Then $\ph_{u,v}(0)=0$ and one computes that $ \|\ph_{u,v}\|_{\Lip}<
\frac{4\pi}{n} $. By the fact that $T^*$ maps the unit ball of
$\ell_\infty^N$ onto the unit ball of $\Lip_0$, it follows that
there is $\phi_{u,v}\in\ell_\infty^N$ with $\|\phi_{u,v}\|_\infty\le
\frac{4\pi}{n}$ and $T^*\phi_{u,v}=\ph_{u,v}$. Now, the functions
$|I(\phi_{u,v})|\in L_1(\sigma)$ are point-wise bounded by the
constant $\frac{4\pi}{n}$, so by Lemma~\ref{lem:lattice} there
exists $x\in \ell_2(X)$ of norm at most $\frac{4\pi}{n}\|R\|\le
16\pi L$ such that $|R(I(\phi_{u,v}))|$ is bounded pointwise by $x$.
But,

\begin{eqnarray*}R\circ I (\phi_{u,v})(s,t)&=&
\F\circ S\circ \Id\circ
T^*(\phi_{u,v})(s,t)\\&=&\F(\ph_{u,v})(s,t)\\&=&
\frac{1}{n^2}\sum_{(k,\ell)\in X} \frac{1}{u+v}\cdot
\sin\left(\frac{2\pi uk}{n}\right)\cdot \sin\left(\frac{2\pi
v\ell}{n}\right)\cdot\sin\left(\frac{2\pi sk}{n}\right)\cdot
\sin\left(\frac{2\pi t\ell}{n}\right)\\
&=& \left\{\begin{array}{ll} \frac{1}{n^2}\cdot \frac{1}{u+v}\cdot
\|(u+v)\ph_{u,v}\|_{\ell_2(X)}^2& (s,t)=(u,v),\\
0 & (s,t)\neq (u,v).\end{array}\right.
\end{eqnarray*}
Observe that
$$
\|(u+v)\ph_{u,v}\|_{\ell_2(X)}^2= \sum_{(k,\ell)\in X}
\sin^2\left(\frac{2\pi uk}{n}\right)\cdot \sin^2\left(\frac{2\pi
v\ell}{n}\right)=\frac{n^2}{4}.
$$
So,
$$
R\circ I (\phi_{u,v})(s,t)=
\left\{\begin{array}{ll} \frac{1}{4(u+v)}& (s,t)=(u,v),\\
0 & (s,t)\neq (u,v).\end{array}\right.
$$
But
\begin{eqnarray*}
(16\pi L)^2\ge \|x\|_2^2\ge \sum_{u,v=1}^n x_{u,v}^2\ge
\sum_{u,v=1}^n \Big[R\circ I
(\phi_{u,v})(u,v)\Big]^2=\frac{1}{16}\sum_{u,v=1}^{n-1}
\frac{1}{(u+v)^2}\ge \frac{\log n}{32},
\end{eqnarray*}
where the last bound follows from comparison with the appropriate
integrals. The proof of Theorem~\ref{thm:main} is complete.
\end{proof}

\subsection{Discretization and minimum weight matching}

In this section we deduce Theorem~\ref{thm:discretization} from
Theorem~\ref{thm:main}. The main tool is the following theorem of
Bourgain~\cite{Bourgain87}, which gives a quantitative version of
Ribe's theorem~\cite{Ribe76}.

\begin{theorem}[Bourgain's quantitative version of Ribe's
theorem~\cite{Bourgain87}]\label{thm:bourgain-ribe} There exists a
universal constant $C$ with the following property. Let $Y$ and $Z$
be Banach spaces, $\dim(Y)=d$. Assume that $\mathscr Y$ is an
$\e$-net in the unit ball of $Y$, $f:\mathscr Y\to Z$ satisfies
$\dist (f)\le D$, and that $\log \log \frac{1}{\e}\ge Cd\log D$.
Then there exists an invertible linear operator $T: Y\to Z$
satisfying $\|T\|\cdot\|T^{-1}\|\le C\cdot D$.
\end{theorem}

\begin{proof}[Proof of Theorem~\ref{thm:discretization}.] Observe that for every $\mu\in \M$, the  measure
$\frac{1}{\mu^+(X)}\cdot\left(\mu^+\otimes\mu^-\right)$ is in
$\Pi(\mu^+,\mu^-)$. Thus
$$
\|\mu\|_\tau\le \frac{1}{\mu^+(X)}\int_{X\times X}
\|x-y\|_2d\mu^+(x)d\mu^-(y)\le \sqrt{2}\cdot(n-1)\cdot\mu^+(X)\le
2n\cdot |\supp(\mu^+)|\cdot \|\mu\|_\infty\le 2n^3\|\mu\|_\infty.
$$
On the other hand, as we have seen in the proof of Lemma~\ref{lem:M
in Prob}, for every $\mu\in \M$, $\|\mu\|_\infty\le \|\mu\|_\tau$.
It follows from these consideration, and Theorems~\ref{thm:main}
and~\ref{thm:bourgain-ribe}, that for every integer $N\ge
e^{e^{C'n^2\log\log n}}$, the set of probability measures $\mathscr
Y\subseteq \Prob_X$ consisting of measures $\mu\in \Prob_X$ such
that for all $x\in X$, $\mu(x)=k/N$ for some $k\in \{0,\ldots,N\}$,
satisfies $c_1(\mathscr Y,\tau)=\Omega\left(\sqrt{\log n}\right)$.
We pass to a family of subsets as follows. Let $M$ be an integer
which will be determined later. For every $\mu\in \mathscr Y$ we
assign a subset $S_\mu\subseteq \{0,\ldots,nM\}^2$ as follows. For
every $(u,v)\in X=\{0,\ldots,n-1\}^2$, if $\mu(u,v)=k/N$, where
$k\in \{0,\ldots, N\}$, then $S_\mu$ will contain arbitrary $k$
distinct points from the set $(uM,vM)+\left\{0,\ldots,\left\lceil
\sqrt{N}\right\rceil\right\}^2$. Provided $M\ge 4\sqrt{N}$, the sets
$\{S_\mu\}_{\mu\in \mathscr Y}$ thus obtained are disjoint $N$ point
subsets of $\{0,\ldots,nM\}^2$, and it is straightforward to check
that the minimum weight matching metric on $\{S_\mu\}_{\mu\in
\mathscr Y}$ is bi-Lipschitz equivalent to $(\mathscr Y,\tau)$ with
constant distortion.
\end{proof}

\subsection{Uniform and coarse nonembeddability into Hilbert space}

In this section we prove Theorem~\ref{thm:negative type}. We shall
prove, in fact, that the space $\M_{[0,1]^2,\tau}$ does not embed
uniformly or coarsely into $L_2$. We first recall the defintions
of these important notions (see~\cite{BL00,MN05} and the
references therein for background on these concepts). Let
$(X,d_X)$ and $(Y,d_Y)$ be metric spaces. For $f:X\to Y$ and $t>0$
we define
$$
\Omega_f(t)=\sup\{d_Y(f(x),f(y));\ d_X(x,y)\le t\},
$$
and
$$
\omega_f(t)=\inf\{d_Y(f(x),f(y));\ d_N(x,y)\ge t\}.
$$
Clearly $\Omega_f$ and $\omega_f$ are non-decreasing, and for
every $x,y\in X$,
$$
\omega_f\left(d_X(x,y)\right)\le d_Y(f(x),f(y))\le
\Omega_f\left(d_X(x,y)\right).
$$
 With these
definitions, $f$ is uniformly continuous if $\lim_{t\to
0}\Omega_f(t)=0$, and $f$ is said to be a uniform embedding if $f$
is invertible and both $f$ and $f^{-1}$ are uniformly continuous.
Also, $f$ is said to be a coarse embedding if $\Omega_f(t)<\infty$
for all $t>0$ and $\lim_{t\to \infty} \omega_f(t)=\infty$.

In what follows we will use the following standard notation: Given a
sequence of Banach spaces
$\left\{(Z_j,\|\cdot\|_{Z_j})\right\}_{j=1}^\infty$ the Banach space
$\left(\bigoplus_{j=1}^\infty Z_j\right)_1$ is the space of all
sequences $\overline z=(z_j)_{j=1}^\infty\in \prod_{j=1}^\infty Z_j$
such that $\|\overline z\|\coloneqq
\sum_{j=1}^\infty\|z_j\|_{Z_j}<\infty$. If for every $j\in \mathbb
N$, $Z_j=Z_1$, we write $\ell_1(Z_1)=\left(\bigoplus_{j=1}^\infty
Z_j\right)_1$.

\begin{theorem}\label{prop:nonUnifEmb}
 The spaces $\left\{\mathscr
M_{\{0,\dots,n\}^2,\tau}^0\right\}_{n=1}^\infty$ do not admit a
uniform or coarse embedding into $L_2$ with moduli uniformly
bounded in $n$, i.e., there do not exist increasing functions
$\omega, \Omega:[0,\infty)\to[0,\infty)$ which either satisfy
$\lim_{t\to 0}\omega(t)=\lim_{t\to 0}\Omega(t)=0$, or
$\lim_{t\to\infty}\omega(t)=\infty$, and mappings $f_n:\mathscr
M_{\{0,\dots,n\}^2}^0\to L_2$, such that
$\omega(\|\mu-\nu\|_\tau)\le \|f_n(\mu)-f_n(\nu)\|_2\le
\Omega(\|\mu-\nu\|_\tau)$ for all $\mu, \nu\in \mathscr
M_{\{0,\dots,n\}^2}^0$ and all $n$.
\end{theorem}

\begin{proof}
If this is not the case then by passing to a limit along an
ultrafilter we easily deduce that $\mathscr M_{[0,1]^2,\tau}^0$
uniformly or coarsely embeds in an ultraproduct of Hilbert spaces
and thus in $L_2$ (see~\cite{Hein80,HM82}). By a theorem of
Aharoni, Maurey and Mityagin~\cite{AMM85} in the case of uniform
embeddings, and a result of Randrianarivony~\cite{Ran04} in the
case of coarse embeddings, this implies that $\mathscr
M_{[0,1]^2}^0$ is linearly isomorphic to a subspace of $L_0$. By a
theorem of Niki\v{s}in~\cite{Nikisin72} it follows that $\mathscr
M_{[0,1]^2}^0$ is isomorphic to a subspace of $L_{1-\e}$ for any
$\e\in (0,1)$. We recall that it is an open problem posed by
Kwapien (see the discussion in~\cite{Kalton85,BL00}) whether a
Banach space which linearly embed into $L_0$ is linearly
isomorphic to a subspace of $L_1$. If this were the case, we would
have finished by Theorem~\ref{thm:main}. Since the solution of
Kwapien's problem is unknown, we proceed as follows.

Let $\{S_j\}_{j=1}^\infty$ be a sequence of disjoint squares in
$[0,1]^2$ with
\begin{eqnarray}\label{eq:diam}d(S_j,S_k)=\min_{a\in
S_j,\ b\in S_k}\|a-b\|_2>\max\left\{{\rm diam}{S_j},{\rm
diam}{S_k}\right\}.\end{eqnarray} Consider the linear subspace $Y$
of $\mathscr M_{[0,1]^2}^0$ consisting of all measures $\mu$
satisfying ${\rm supp}(\mu)\subseteq\bigcup_{j=1}^\infty S_j$ and
$\mu(S_j)=0$ for all $j$. It is intuitively clear that in the
computation of $\|\mu\|_\tau$ for $\mu\in Y$ the best
transportation leaves each of the $S_j$ invariant; i.e., it is
enough to take the infimum in~\eqref{eq:def tau} only over
measures $\pi\in\Pi(\mu,\nu)$ which are supported on
$\bigcup_{j=1}^\infty (S_j\times S_j)$. This is proved formally as
follows: Fix $\mu\in Y$ and write $\mu=\sum_{j=1}^\infty\mu_j$,
where $\supp(\mu_j)\subseteq S_j$ and $\mu_j(S_j)=0$ for all $j\in
\mathbb N $. We claim that
\begin{eqnarray}\label{eq:l1 sum}
\|\mu\|_{[0,1]^2,\tau}=\sum_{j=1}^\infty\|\mu_j\|_{S_j,\tau}.
\end{eqnarray}
If $\pi_j\in \Pi(\mu_j^+,\mu_j^-)$ then $\pi\coloneqq
\sum_{j=1}^\infty \pi_j\in \Pi(\mu^+,\mu^-)$. Thus
$\|\mu\|_{[0,1]^2,\tau}\le\sum_{j=1}^\infty\|\mu_j\|_{S_j,\tau}$. To
prove the reverse inequality take $\pi\in \Pi(\mu^+,\mu^-)$. For
every $j=1,2,\ldots$ define a measure $\sigma_j$ on $S_j$ as
follows: For $A\subseteq S_j$ set $ \sigma_j(A)\coloneqq
\pi\left(A\times \bigcup_{k\neq j} S_k\right)$. Thus, in particular,
by our assumption~\eqref{eq:diam} for every $y\in S_j$,
\begin{eqnarray}\label{eq:use diam}
\int_{S_j}\|x-y\|_2d\sigma_j(x)=\int_{S_j\times \bigcup_{k\neq
j}S_k}\|x-y\|_2d\pi(x,z)\le\int_{S_j\times \bigcup_{k\neq
j}S_k}\|x-z\|_2d\pi(x,z).
\end{eqnarray}
Writing
$$
\widetilde \pi\coloneqq \pi\cdot {\bf 1}_{\bigcup_{j=1}^\infty
(S_j\times S_j)}+\sum_{j=1}^\infty\frac{1}{\sigma_j(S_j)}\cdot
\sigma_j\otimes\sigma_j=\pi\cdot {\bf 1}_{\bigcup_{j=1}^\infty
(S_j\times S_j)}+\sum_{j=1}^\infty\frac{1}{\pi\left(S_j\times
\bigcup_{k\neq j} S_k\right)}\cdot \sigma_j\otimes\sigma_j,
$$
it follows from our definitions that $\widetilde \pi \in
\Pi(\mu^+,\mu^-)$ and $\widetilde \pi$ is supported on
$\bigcup_{j=1}^\infty (S_j\times S_j)$. Moreover, for each $j$,
$\widetilde \pi_j\coloneqq\widetilde \pi|_{S_j}\in
\Pi(\mu_j^+,\mu_j^-)$, so that
\begin{eqnarray*}\label{eq:stup step}
\sum_{j=1}^\infty\|\mu_j\|_{S_j,\tau}&\le&\nonumber
\sum_{j=1}^\infty \int_{S_j\times S_j}
\|x-y\|_2d\widetilde\pi_j(x,y)\\&=&\int_{\bigcup_{j=1}^\infty
(S_j\times
S_j)}\|x-y\|_2d\pi(x,y)+\sum_{j=1}^\infty\frac{1}{\pi\left(S_j\times
\bigcup_{k\neq j} S_k\right)}\cdot\int_{S_j\times S_j}
\|x-y\|_2d\sigma_j(x)d\sigma_j(y)\\
&\stackrel{\eqref{eq:use diam}}{\le}& \int_{\bigcup_{j=1}^\infty
(S_j\times S_j)}\|x-y\|_2d\pi(x,y)+\sum_{j=1}^\infty\int_{S_j\times
\bigcup_{k\neq j}S_k}\|x-z\|_2d\pi(x,z)\\
&=& \int_{\left(\bigcup_{j=1}^\infty S_j\right)\times
\left(\bigcup_{j=1}^\infty S_j\right)}\|x-y\|_2d\pi(x,y).
\end{eqnarray*}
This concludes the proof of~\eqref{eq:l1 sum}. It follows that $Y$
is isometric to $\left(\bigoplus_{n=1}^\infty \mathscr
M_{S_n,\tau}^0\right)_1$, which in turn is isometric to
$\ell_1\left( \mathscr M_{[0,1]^2,\tau}^0\right)$. Now, Kalton
proved in~\cite{Kalton85} that if for some Banach space $X$,
$\ell_1(X)$ is isomorphic to a subspace of $L_{0}$, then $X$ is
isomorphic to a subspace of $L_1$ and we finish by
Theorem~\ref{thm:main}.
\end{proof}

\begin{proof}[Proof of Theorem~\ref{thm:negative type}] Assume for
the sake of contradiction that there exists $C<\infty$ such that
for all $n\in \mathbb N$,
$c_2\left(\Prob_{\{0,\ldots,n\}^2},\sqrt{\tau}\right)< C$. By the
proof of Lemma~\ref{lem:M in Prob} we know that the unit ball of
$\M_{\{0,\ldots,n\}^2,\tau}$ is isometric to a subset of
$(\Prob_{\{0,\ldots,n\}^2},\tau)$. Thus by our assumption there
exist mappings $f_n:\M_{\{0,\ldots,n\}^2}\to L_2$ such that for
every $\mu,\nu\in \M_{\{0,\ldots,n\}^2}$ with
$\|\mu\|_\tau,\|\nu\|_\tau\le 1$,
\begin{eqnarray}\label{eq:sqrt}
\sqrt{\|\mu-\nu\|_\tau}\le \|f_n(\mu)-f_n(\nu)\|_2\le
C\cdot\sqrt{\|\mu-\nu\|_\tau}\enspace .
\end{eqnarray}
Let $\mathscr U$ be a free ultrafilter on $\mathbb N$. Define
$\widetilde f_n: \M_{\{0,\ldots,n\}^2}\to (L_2)_{\mathscr U}$ by
$\widetilde f_n(\mu)=\left(\sqrt{j}\cdot
f_n(\mu/j)\right)_{j=1}^\infty/\mathscr U$.
Inequalities~\eqref{eq:sqrt} imply that all $\mu,\nu\in
\M_{\{0,\ldots,n\}^2}$ satisfy $\sqrt{\|\mu-\nu\|_\tau}\le
\|\widetilde f_n(\mu)-\widetilde f_n(\nu)\|_{(L_2)_\mathscr U}\le
C\cdot\sqrt{\|\mu-\nu\|_\tau}$. Since the ultrapower
$(L_2)_\mathscr U$ is isometric to a Hilbert space
(see~\cite{Hein80}), we arrive at a contradiction with
Theorem~\ref{prop:nonUnifEmb}.
\end{proof}

\begin{remark}\label{rem:quant}{\em
We believe that Theorem~\ref{thm:negative type} can be made
quantitative, i.e. one can give explicit quantitative estimates on
the rate with which
$c_2\left(\Prob_{\{0,\ldots,n\}^2},\sqrt{\tau}\right)$ tends to
infinity. This would involve obtaining quantitative versions of
the proofs in~\cite{AMM85,Kalton85,Ran04}, which seems easy but
somewhat tedious. We did not attempt to obtain such bounds.}
\end{remark}

\begin{remark}\label{rem:ball}{\em We do not know whether $\left(\Prob_{[0,1]^2},\tau\right)$
admits a uniform embedding into Hilbert space. The proof above
actually gives that for all $\alpha\in (0,1]$,
$\left(\Prob_{[0,1]^2,\tau},\tau^\alpha\right)$ does not embed
bi-Lipschitzly into Hilbert space. But, our proof exploits the
homogeneity of the function $t\mapsto t^\alpha$ in an essential
way, so it does not apply to the case of more general moduli. }
\end{remark}

\section{Upper bounds via Fourier analysis}\label{sec:upper}

In this section we prove Theorem~\ref{thm:upper}, and discuss some
related upper bounds. Given a measure $\mu$ on $\Z_n^2$ we
decompose it as
 in~\eqref{eq:decompose}, and we consider the linear operators $A$
 and $B$, from $\M_{\Z_n^2}$ to $L_1\left(\Z_n^2\right)$, defined in~\eqref{eq:def A} and~\eqref{eq:def B},
 respectively. One checks that the duals of these operators,
 $A^*,B^*:L_1\left(\Z_n^2\right)\to
 \M_{\Z_n^2}^*=\Lip_0\left(\Z_n^2\right)$, are given by
\begin{eqnarray}\label{eq:def A^*}
 A^*f=\sum_{(u,v)\in \Z_n^2\setminus \{(0,0)\}} \frac{e^{-\frac{2\pi
i u}{n}}-1}{\big|e^{\frac{2\pi i u}{n}}-1\big|^2+\big|e^{\frac{2\pi
i v}{n}}-1\big|^2}\cdot \widehat f(u,v)\cdot (e_{uv}-1),
\end{eqnarray}
and
\begin{eqnarray}\label{eq:def B^*}
B^*f=\sum_{(u,v)\in \Z_n^2\setminus \{(0,0)\}} \frac{e^{-\frac{2\pi
i v}{n}}-1}{\big|e^{\frac{2\pi i u}{n}}-1\big|^2+\big|e^{\frac{2\pi
i v}{n}}-1\big|^2}\cdot \widehat f(u,v)\cdot (e_{uv}-1).
\end{eqnarray}
To check these identities the reader should verify that for all
$\mu\in \M_{\Z_n^2}$, $\int_{\Z_n^2} fd (A\mu)= \int_{\Z_n^2}
(A^*f)d\mu$, and similarly for $B$ (to this end, recall that
$\mu\left(\Z_n^2\right)=0$, so that $\widehat \mu(0,0)=0$. This
explains the subtraction of $1$ in the identities~\eqref{eq:def A^*}
and~\eqref{eq:def B^*}).

 We claim that for every $\mu\in
 \M_{Z_n^2}$,
\begin{eqnarray}\label{eq:goal} \|\mu\|_\tau\le
\|A\mu\|_{L_1\left(\Z_n^2\right)}+\|B\mu\|_{L_1\left(\Z_n^2\right)}\le
C\log n\cdot \|\mu\|_\tau,
 \end{eqnarray}
where $C$ is a universal constant. This will imply
Theorem~\ref{thm:upper} since the mapping $\mu\mapsto \mu-U$, where
$U$ is the uniform probability measure on $Z_n^2$, is an isometric
embedding of $\Prob_{\Z_n^2}$ into $\M_{\Z_2^n}$.

By duality, \eqref{eq:goal} is equivalent to the fact that the
mapping $(f,g)\mapsto A^*f+B^*g$ from
$L_\infty\left(\Z_n^2\right)\oplus L_\infty\left(\Z_n^2\right)$ to
$\Lip_0\left(\Z_n^2\right)$ is a $C\log n$ quotient map, i.e. for
every $(f,g)\in L_\infty\left(\Z_n^2\right)\oplus
L_\infty\left(\Z_n^2\right)$
\begin{eqnarray}\label{eq:dual goal}
\left\|A^*f+B^*g\right\|_{\Lip} \le C\log n \cdot\max
\left\{\|f\|_\infty,\|g\|_\infty\right\},
\end{eqnarray}
and for every $h\in \Lip_0\left(\Z_n^2\right)$ there is some
$(f,g)\in L_\infty\left(\Z_n^2\right)\oplus
L_\infty\left(\Z_n^2\right)$ satisfying $A^*f+B^*g= h$ and
$\max\{\|f\|_\infty, \|g\|_\infty\}\le \|h\|_{\Lip}$. The second
assertion is proved as follows: Take $f=\partial_1 h$ and
$g=\partial_2 h$, where for $j=1,2$, $\partial_j
h(x)=h(x+e_j)-h(x)$ (here $e_1=(1,0)$ and $e_2=(0,1)$). Clearly
$\|f\|_\infty,\|g\|_\infty\le \|h\|_{\Lip}$, and
\begin{eqnarray*}
A^*f+B^*g&=& \sum_{(u,v)\in \Z_n^2\setminus \{(0,0)\}}\left(
\frac{\left(e^{-\frac{2\pi i u}{n}}-1\right)\cdot
\widehat{\partial_1 h}(u,v)+\left(e^{-\frac{2\pi i
v}{n}}-1\right)\cdot  \widehat{\partial_2 h}(u,v)
}{\big|e^{\frac{2\pi i u}{n}}-1\big|^2+\big|e^{\frac{2\pi i
v}{n}}-1\big|^2}\right)(e_{uv}-1)\\
&=& \sum_{(u,v)\in \Z_n^2\setminus \{(0,0)\}}\left(
\frac{\left(e^{-\frac{2\pi i u}{n}}-1\right)\cdot
\left(e^{\frac{2\pi i u}{n}}-1\right)+\left(e^{-\frac{2\pi i
v}{n}}-1\right)\cdot \left(e^{\frac{2\pi i
v}{n}}-1\right)}{\big|e^{\frac{2\pi i
u}{n}}-1\big|^2+\big|e^{\frac{2\pi i v}{n}}-1\big|^2}\right)\cdot
\widehat h(u,v)(e_{uv}-1)\\
&=& \sum_{(u,v)\in \Z_n^2\setminus \{(0,0)\}}\widehat h(u,v)e_{uv}-
\sum_{(u,v)\in \Z_n^2\setminus \{(0,0)\}}\widehat h(u,v)\\
&=& \sum_{(u,v)\in \Z_n^2}\widehat h(u,v)e_{uv}=h,
\end{eqnarray*}
where we used the fact that $h(0)=0$.

It remains to prove~\eqref{eq:dual goal}. To this end, it is enough
to show that $\|A^* f\|_{\Lip}\le O(\log n) \cdot \|f\|_\infty$ and
$\|B^* g\|_{\Lip}\le O(\log n) \cdot \|g\|_\infty$. We will
establish this for $A^*$- the case of $B^*$ is entirely analogous.
Observe that
$$
\|A^* f\|_{\Lip}\le \|\partial_1 A^* f\|_\infty+\|\partial_2 A^*
f\|_\infty,
$$
so it is enough to establish the following two inequalities:
\begin{eqnarray}\label{eq:before p1}
\left\|\sum_{(u,v)\in \Z_n^2\setminus \{(0,0)\}}
\frac{\big|e^{\frac{2\pi i u}{n}}-1\big|^2}{\big|e^{\frac{2\pi i
u}{n}}-1\big|^2+\big|e^{\frac{2\pi i v}{n}}-1\big|^2}\cdot \widehat
f(u,v)e_{uv}\right\|_\infty\le O(\log n)\cdot \|f\|_\infty,
\end{eqnarray}
and
\begin{eqnarray}\label{eq:before p2}
\left\|\sum_{(u,v)\in \Z_n^2\setminus \{(0,0)\}}
\frac{\left(e^{-\frac{2\pi i u}{n}}-1\right)\cdot\left(e^{\frac{2\pi
i v}{n}}-1\right)}{\big|e^{\frac{2\pi i
u}{n}}-1\big|^2+\big|e^{\frac{2\pi i v}{n}}-1\big|^2}\cdot \widehat
f(u,v)e_{uv}\right\|_\infty\le O(\log n)\cdot \|f\|_\infty.
\end{eqnarray}
Since for $p>0$ the norms on $L_\infty\left(\Z_n^2\right)$ and
$L_p\left(\Z_n^2\right)$ are equivalent with constant $n^{2/p}$ (by
H\"older's inequality), it is enough to show that for $p\ge 2$,
\begin{eqnarray}\label{eq:after p1}
\left\|\sum_{(u,v)\in \Z_n^2\setminus \{(0,0)\}}
\frac{\big|e^{\frac{2\pi i u}{n}}-1\big|^2}{\big|e^{\frac{2\pi i
u}{n}}-1\big|^2+\big|e^{\frac{2\pi i v}{n}}-1\big|^2}\cdot \widehat
f(u,v)e_{uv}\right\|_p\le O(p)\cdot \|f\|_p,
\end{eqnarray}
and
\begin{eqnarray}\label{eq:after p2}
\left\|\sum_{(u,v)\in \Z_n^2\setminus \{(0,0)\}}
\frac{\left(e^{-\frac{2\pi i u}{n}}-1\right)\cdot\left(e^{\frac{2\pi
i v}{n}}-1\right)}{\big|e^{\frac{2\pi i
u}{n}}-1\big|^2+\big|e^{\frac{2\pi i v}{n}}-1\big|^2}\cdot \widehat
f(u,v)e_{uv}\right\|_p\le O(p)\cdot \|f\|_p.
\end{eqnarray}

To prove inequalities~\eqref{eq:after p1} and~\eqref{eq:after p2}
we will assume that $n$ is odd (all of our results are valid for
even $n$ as well, and the proofs in this case require minor
modifications). We think of $\Z_n^2$ as $[-(n-1)/2,(n-1)/2]\cap
\Z$. As before, given $m:\Z_n^2\to \C$ we denote
$$
\partial_1 m(x,y)=m(x+1,y)-m(x,y),\quad \mathrm{and}\quad
\partial_2 m(x,y)=m(x,y+1)-m(x,y).
$$
Thus
$$
\partial_1^2 m(x,y)=m(x+2,y)-2m(x+1,y)+m(x,y)\quad
\mathrm{and}\quad \partial_2^2 m(x,y)=m(x,y+2)-2m(x,y+1)+m(x,y),
$$
and
$$
\partial_1\partial_2 m(x,y)=\partial_2\partial_1
m(x,y)=m(x+1,y+1)-m(x+1,y)-m(x,y+1)-m(x,y).
$$

In what follows we think of $m$ as a {\em Fourier multiplier} in
the sense that it corresponds to a translation invariant operator
$T_m$ on $L_2\left(\Z_n^2\right)$ given by
\begin{eqnarray}\label{eq;def multiplier}
T_m(f)\coloneqq  \sum_{(u,v)\in \Z_n^2} m(u,v)\cdot \widehat
f(u,v) \cdot e_{uv}.
\end{eqnarray}

Recall that an operator $T:L_1\left(\Z_n^2\right)\to
L_1\left(\Z_n^2\right)$ is said to be weak $(1,1)$ with constant
$K$ if for every $f:\Z_n^2\to \C$ and every $a>0$,
$$
\left|\left\{((u,v)\in \Z_n^2:\ |Tf(u,v)|\ge a \right\}\right|\le
\frac{K}{a}\cdot \|f\|_1= \frac{K}{a}\cdot\sum_{(u,v)\in
\Z_n^2}|f(u,v)|.
$$

We will use the following discrete version of the
H\"ormander-Mihlin multiplier theorem~\cite{Mih56,Hor60}.

\begin{theorem}[H\"ormander-Mihlin multiplier criterion on $\Z_n^2$]\label{thm:hormander} For $j\in \mathbb N$ denote
$Q_j=[-2^j,2^j]\times [-2^j,2^j]$. Fix $B>0$ and $m:\Z_n^2\to \C$
with $m(0,0)=0$, and assume that for all
$j=0,1,\ldots,\lfloor\log_2 (n-1)\rfloor-1$,
\begin{eqnarray*}
\sum_{(u,v)\in (Q_j\setminus Q_{j-1})\cap
\Z_n^2}&&\!\!\!\!\!\!\!\!\!\!\!\!\!
\left[2^{-2j}|m(u,v)|^2+|\partial_1 m(u,v)|^2+|\partial_2
m(u,v)|^2+\right.\\&\phantom{\le}&\quad\left.2^{2j}|\partial_1^2
m(u,v)|^2+ 2^{2j}|\partial_2^2
m(u,v)|^2+2^{2j}|\partial_1\partial_2 m(u,v)|^2\right]\le B^2.
\end{eqnarray*}
Then the translation invariant operator $T_m$ corresponding to $m$
is weak $(1,1)$ with constant $O(B)$.
\end{theorem}

While the continuous version of the H\"ormander-Mihlin multiplier
theorem is a powerful tool which appears in several texts (e.g. in
the books~\cite{G-CRdeF85,Stein93,Tor04}), we could not locate a
statement of the above discrete version in the literature. It is,
however, possible to prove it using several minor modifications of
the existing proofs. The standard proof of the H\"ormander-Mihlin
criterion is usually split into two parts. The first part, which
is based on the Calder\'on-Zygmund decomposition, transfers
virtually verbatim to the discrete setting- see Theorem 3 in
Chapter 1 of~\cite{Stein93}, and Remark 8.1 there which explains
how this part of the proof transfers from $\R^n$ to the setting of
finitely generated groups of polynomial growth (in fact, the
Calder\'on-Zygmund decomposition itself, as presented in Theorem 2
in Chapter 1 of~\cite{Stein93}, is valid in the setting of general
metric spaces equipped with a doubling measure). The second part
of the proof of the H\"ormander-Mihlin theorem, as presented in
Theorem 2.5 of~\cite{Hor60}, requires several straightforward
modifications in order to pass to the discrete setting. We leave
the simple details to the reader. For the sake of readers that are
not familiar with these aspects of Fourier analysis, we will later
present a complete reduction to a continuous problem whose proof
appears in print, which yields slightly worse bounds on the
distortion guarantee.

In order to apply Theorem~\ref{thm:hormander} we consider the
following two multipliers,
\begin{eqnarray}\label{eq:ms}
m_1(u,v)\coloneqq \frac{\big|e^{\frac{2\pi i
u}{n}}-1\big|^2}{\big|e^{\frac{2\pi i
u}{n}}-1\big|^2+\big|e^{\frac{2\pi i v}{n}}-1\big|^2}, \quad
\mathrm{and}\quad m_2(u,v)\coloneqq \frac{\left(e^{-\frac{2\pi i
u}{n}}-1\right)\cdot\left(e^{\frac{2\pi i
v}{n}}-1\right)}{\big|e^{\frac{2\pi i
u}{n}}-1\big|^2+\big|e^{\frac{2\pi i v}{n}}-1\big|^2},
\end{eqnarray}
where we set $m_1(0,0)=m_2(0,0)=0$. A direct (albeit tedious!)
computation shows that $m_1$ and $m_2$ satisfy the conditions of
Theorem~\ref{thm:hormander} with $B=O(1)$. Thus, the operators
$T_{m_1}$ and $T_{m_2}$ are weak $(1,1)$ with constant $O(1)$.
Since $m_1$ and $m_2$ are bounded functions, the operator norms
$\|T_{m_1}\|_{L_2\left(\Z_n^2\right)\to L_2\left(\Z_n^2\right)}$
and $\|T_{m_2}\|_{L_2\left(\Z_n^2\right)\to
L_2\left(\Z_n^2\right)}$ are $O(1)$. Since these operators are
self adjoint, by the Marcinkiewicz interpolation theorem
(see~\cite{Zyg02}) it follows that for $p\ge 2$, the operator
norms $\|T_{m_1}\|_{L_p\left(\Z_n^2\right)\to
L_p\left(\Z_n^2\right)}$ and
$\|T_{m_2}\|_{L_p\left(\Z_n^2\right)\to L_p\left(\Z_n^2\right)}$
are $O(p)$. This is precisely~\eqref{eq:after p1}
and~\eqref{eq:after p2}.

\bigskip The above argument is based on
Theorem~\ref{thm:hormander}, which does not appear exactly as
stated in the literature, but its proof is a straightforward
adaptation of existing proofs (which is too simple to justify
rewriting the lengthy argument here). However, making the
necessary changes easily does require some familiarity with
Calder\'on-Zygmund theory. We therefore present now another
argument which gives a $\polylog (n)$ bound on the distortion, but
uses only statements which appear in the literature. This
alternative approach appears to be quite versatile, and might be
useful elsewhere.

The following lemma reduces the problem of proving inequalities
such as~\eqref{eq:after p1} and~\eqref{eq:after p2} (with perhaps
a different dependence on $p$) to a continuous inequality. The
argument is based on the proof of a theorem of Marcinkiewicz
from~\cite{Zyg02} (see Theorem 7.5 in chapter X there). In what
follows we denote by $\mathbb T$ the Euclidean unit circle in the
plane.

\bigskip

\begin{prop}[Transferring multipliers from the torus to $\mathbb
Z_n^2$]\label{prop:transfer} Fix an odd integer $n$. Let
$\{\lambda(u,v)\}_{u,v=0}^\infty$ be complex numbers such that
$\lambda(u,v)=0$ for $\max\{u,v\}\ge n$. Consider the operators
$M:L_p\left(\mathbb T^2\right)\to L_p\left(\mathbb T^2\right)$ and
$M_n:L_p\left(\mathbb Z_n^2\right)\to L_p\left(\mathbb
Z_n^2\right)$ given by
\[
M\left(\sum_{u,v=-\infty}^\infty\widehat f(u,v)e^{2\pi i(
ux+vy)}\right)=\sum_{u,v=0}^\infty \lambda(u,v)\widehat
f(u,v)e^{2\pi i (ux+vy)},
\]
and
\[
M_n\left(\sum_{u,v=0}^{n-1}\widehat f(u,v)e^{\frac{2\pi i}{n}(
ua+vb)}\right)=\sum_{u,v=0}^{n-1}\lambda(u,v)\widehat
f(u,v)e^{\frac{2\pi i}{n}( ua+vb)}.
\]
Then, $$\|M_n\|_{L_p\left(\mathbb Z_n^2\right)\to L_p\left(\mathbb
Z_n^2\right)}\le 81\cdot\|M\|_{L_p(\mathbb T^2)\to L_p(\mathbb
T^2)}.$$
\end{prop}

\begin{proof} The proof is a variant of the first part of the proof of Theorem 7.5 in chapter X in
\cite{Zyg02}, and a small twist on the second part. Since the
terminology in \cite{Zyg02} is different from ours, we repeat the
proof of the first part as well.  Recall that the Dirichlet kernels
$D_\ell:[0,1]\to \mathbb C$ are defined as,
\[
D_\ell (x)=\sum_{j=-\ell}^{\ell}e^{2\pi i jx},
\]
and the Fej\' er kernels $K_m:[0,1]\to \mathbb C$ are
\[
K_m (x) =\frac1{m+1}\sum_{\ell=0}^m D_\ell
x=\sum_{j=-m}^m\left(1-\frac{|j|}{m+1}\right)e^{2\pi ijx}.
\]
A basic property of  $D_\ell$  is that for any trigonometric
polynomial $S(x)$ of degree at most $\ell$, namely
$S(x)=\sum_{j=-\ell}^{\ell}a_je^{2\pi i jx}$, we have that \
$S(x)=S* D_\ell(x)=\int_0^1 S(t)D_\ell(x-t)dt$. The same is true
with any other function all of whose $j$th Fourier coefficients for
$j$ between $-\ell$ and $\ell$ are  $1$; in particular for the de la
Vall\'ee Poussin kernel $2K_{2\ell-1}-K_{\ell-1}$
(see~\cite{Katz04}). The well known advantage of the Fej\'er kernel
over the Dirichlet kernel is that it is everywhere (real and)
nonnegative. Note also that $\int_0^1K_m(t)dt=1$ for all $m$. Thus,
by convexity of the function $t^p$, for any trigonometric polynomial
$S$ of degree at most $\ell$, and for all $x\in[0,1]$,
\begin{eqnarray}\label{eq:conv}
|S(x)|^p&= & |2S* K_{2\ell-1}(x)-S* K_{\ell}(x)|^p\nonumber\\
&\le&  3^p\left(\frac23\int_0^1|S(t)|^pK_{2\ell-1}(x-t)dt+
\frac13\int_0^1|S(t)|^pK_{\ell-1}(x-t)dt\right).
\end{eqnarray}

Let now $\omega_{2\ell+1}$ be the measure which assign mass
$\frac{1}{2\ell+1}$ to each of $2\ell+1$ equally spaced points on
$[0,1]$. Then it is easy to check that
\[
\int_0^1K_m(x-t)d\omega_{2\ell+1}(x)=\int_0^1K_m(x-t)dx=1
\]
for all $m\le 2\ell$ and for all $t\in [0,1]$. Integrating
(\ref{eq:conv}) with respect to $\omega_{2\ell+1}$, we get that for
any trigonometric polynomial $S$ of degree at most $\ell$
\begin{equation}\label{eq:upperbnd}
\int_0^1|S(x)|^pd\omega_{2\ell+1}(x)\le 3^p\int_0^1|S(x)|^pdx.
\end{equation}
It follows that if $S(x,y)$ is a two-variable trigonometric
polynomial of degree at most $\ell$ in each of the variables, i.e.
$S(x,y)=\sum_{u,v=-\ell}^{\ell}a_{uv}e^{2\pi i (ux+vy)}$,
\begin{equation*}
\int_{[0,1]^2}|S(x,y)|^pd\omega_{2\ell+1}(x)d\omega_{2\ell+1}(y)\le
9^p\int_{[0,1]^2}|S(x,y)|^pdxdy.
\end{equation*}

It follows from this that, since $n$ is odd, for every $f\in
L_p\left(\mathbb T^2\right)$,
\begin{equation*}
\left\|M_n\left(\sum_{u,v=0}^{n-1}\widehat f(u,v)e^{\frac{2\pi
i}{n}( ua+vb)}\right)\right\|_{L_p\left(\mathbb Z_n^2\right)}\le 9
\left\|M\left(\sum_{u,v=-\infty}^\infty\widehat f(u,v)e^{2\pi i(
ux+vy)}\right)\right\|_{L_p\left(\mathbb T^2\right)}.
\end{equation*}

Note that for each trigonometric polynomial of the form
$P(x,y)=\sum_{u,v=-n+1}^{n-1}a_{uv} e^{2\pi i ({ux+vy})}$,
\[
\int_{[0,1]^2} P(x,y) d\omega_n(x)d\omega_n(y)=a_0=\int_{[0,1]^2}
P(x,y) dxdy.
\]
Fix $f\in L_p\left(\mathbb Z_n^2\right)$, $1< p <\infty$. By the
first part of the proof and duality, there is $g\in
L_{p^*}(\mathbb T^2)$ ($p^*=p/(p-1)$) with $\|g\|_{p^*}=1$ such
that
\begin{eqnarray*}
\|M_nf\|_{L_p\left(\mathbb Z_n^2\right)}&\le&  9\int_{[0,1]^2}
\left(\sum_{u,v=0}^{n-1}
\lambda_j\widehat f(u,v)e^{2\pi i (ux+vy)}\right)\overline{ g(x,y)}dxdy\\
&= & 9\int_{[0,1]^2}\left(\sum_{u,v=0}^{n-1} \lambda(u,v)\widehat
f(u,v)e^{2\pi i (ux+vy)}\right)\left(\sum_{u,v=0}^{n-1}
\overline{\widehat g(u,v)}e^{-2\pi i
(ux+vy)}\right)dxdy\\
&= & 9\int_{[0,1]^2} \left(\sum_{u,v=0}^{n-1} \lambda(u,v)\widehat
f(u,v)e^{2\pi i (ux+vy)}\right)\left(\sum_{u,v=0}^{n-1}
\overline{\widehat g(u,v)}e^{-2\pi i
(ux+vy)}\right)d\omega_n(x)d\omega_n(y)\\
&= & 9\int_{[0,1]^2} \left(\sum_{u,v=0}^{n-1} \widehat f(u,v)e^{2\pi
i (ux+vy)}\right)\left(\sum_{u,v=0}^{n-1}
\lambda(u,v)\overline{\widehat g(u,v)}e^{-2\pi i(
ux+vy)}\right)d\omega_n(x)d\omega_n(y)\\
&\le & 9\left(\int_{[0,1]^2} \left|\sum_{u,v=0}^{n-1} \widehat
f(u,v)e^{2\pi i
(ux+vy)}\right|^pd\omega_n(x)d\omega_n(y)\right)^{1/p}\cdot\\&\phantom{\le}&\
\ \ \ \ \ \ \ \ \  \left(\int_{[0,1]^2} \left| \sum_{u,v=0}^{n-1}
\lambda(u,v)\overline{\widehat g(u,v)}e^{-2\pi i
(ux+vy)}\right|^{p^*}d\omega_n(x)\omega_n(y)\right)^{1/p^*}\\
&\le & 81\cdot \|f\|_{L_p\left(\mathbb
Z_n^2\right)}\left(\int_{[0,1]^2} \left| \sum_{u,v=0}^{n-1}
\overline{\lambda(u,v)}\widehat g(u,v)e^{2\pi i
(ux+vy)}\right|^{p^*}dx\right)^{1/p^*}\\
&\le&  81 \cdot\|f\|_{L_p\left(\mathbb
Z_n^2\right)}\cdot\|M\|_{L_p\left(\mathbb T^2\right)\to
L_p\left(\mathbb T^2\right)}.
\end{eqnarray*}
where the inequality before last follows from \eqref{eq:upperbnd}
and the last inequality (that is the fact that the norm of a
multiplier in $L_p\left(\mathbb T^2\right)$ is the same as the norm
of the conjugate multiplier in $L_{p^*}\left(\mathbb T^2\right)$)
follows from duality. The case $p=1$ (and also a similar inequality
for the $\infty$ norm) follows easily from the $L_p$ cases.
\end{proof}

Proposition~\ref{prop:transfer} implies that it is enough to
obtain $L_p$ to $L_p$ bounds for the operators $T_{m_1}$ and
$T_{m_2}$, where $m_1,m_2$ are as in~\eqref{eq:ms}, as operators
on functions on the torus $\mathbb T^2$. By a theorem of de
Leeuw~\cite{deLeeuw65} it is enough to obtain such bounds when we
think of $T_{m_1}$ and $T_{m_2}$ as operators on functions on
$\R^2$ (see~\cite{Woz96} for the respective result in the case of
weak $(1,1)$ bounds). The continuous version of the
H\"ormander-Mihlin multiplier theorem now applies, but
unfortunately its conditions are not satisfied. However, a (once
again tedious) computation shows it is possible to apply the
Marcinkiewicz multiplier theorem (see~\cite{Stein70,Tor04}), in
combination with bounds on the Hilbert
transform~\cite{Stein70,Tor04}, to obtain bounds similar
to~\eqref{eq:after p1} and~\eqref{eq:after p2} with $O(p)$
replaced by $O(\poly(p))$ (it is quite easy to obtain a bound of
$O(p^3)$, and with more work this can be reduced to $O(p^2)$.
However we do not see a simple way to obtain $O(p)$ using this
approach).

\begin{remark}\label{rem:polylog} {\em Consider the mapping
$S:\Prob_{\Z_n^2}\to L_1\left(\Z_n^2\right)$ given by
$$
S\mu\coloneqq \sum_{(u,v)\in \Z_n^2\setminus \{(0,0)\}}
\left(\left|e^{\frac{2\pi i u}{n}}-1\right|+\left|e^{\frac{2\pi i
v}{n}}-1\right|\right)\cdot \widehat \mu(u,v)e_{uv}.
$$
Using considerations similar to the above (see Proposition III.A.3
in~\cite{Woj96} for a continuous counterpart) it is possible to
show that $S$ has distortion $O(\polylog(n))$. However, we were
unable to get this bound down to $O(\log n)$ as in
Theorem~\ref{thm:upper}. Nevertheless, this embedding might be of
interest since it reduces the dimension of the ambient $L_1$ space
by a factor of $2$. }
\end{remark}

\section{Discussion and open problems}

There are several interesting problems that arise from the results
presented in this paper- we shall discuss some of them in the list
below.

\begin{enumerate}

\item The most natural problem is to determine the asymptotic
behavior of $c_1\left(\{0,1\ldots,n\}^2,\tau\right)$. It seems
hard to use the ideas in Section~\ref{sec:upper} to obtain an
embedding of distortion $O\left(\sqrt{\log n}\right)$, as the
known bounds on multipliers usually give a weak $(1,1)$ inequality
at best.

\item Remark~\ref{rem:polylog} implies that the Banach-Mazur
distance between the $n^2-1$ dimensional normed space
$\M_{\Z_n^2,\tau}$ and $\ell_1^{n^2-1}$ is $O(\polylog(n))$. It
would be interesting to determine the asymptotic behavior of this
distance. In particular, it isn't clear whether the $L_1$
(embedding) distortion of $\M_{\Z_n^2,\tau}$ behaves differently
from its Banach-Mazur distance from $\ell_1^{n^2-1}$.

\item We did not attempt to study the $L_1$ distortion of
$\M_{\{0,1,\ldots,n\}^d,\tau}$ for $d\ge 3$. Observe that this
space contains $\M_{\{0,1,\ldots,n\}^2,\tau}$, so the
$\Omega\left(\sqrt{\log n}\right)$ lower bound still applies. But,
the result of~\cite{KN05} shows that the transportation cost
metric on the Hamming cube $\{0,1\}^d$ has distortion $\Theta(d)$,
so some improvements are still possible. Note that in higher
dimensions it becomes interesting to study the transportation cost
distance when $\R^d$ is equipped with other norms. The
Banach-Mazur distance between $\ell_1^d$ and arbitrary
$d$-dimensional norms has been studied
in~\cite{BS88,ST89,Giann95}. In particular, the result
of~\cite{Giann95} states that any $d$-dimensional Banach space is
at distance $O\left(d^{5/6}\right)$ from $\ell_1^d$. Combining
this fact with the lower bound on the $L_1$ distortion of the
transportation cost distance on the Hamming ($\ell_1$) cube cited
above, we see that for any norm $\|\cdot \|$ on $\R^d$,
$c_1\left(\Prob_{(\R^d,\|\cdot\|),\tau}\right)=\Omega\left(d^{1/6}\right)$.
It would be interesting to study the dependence on $d$ for general
norms on $\R^d$.

\item As stated in Remark~\ref{rem:quant}, it would be interesting
to study the rate with which
$c_2\left(\Prob_{\{0,\ldots,n\}^2},\sqrt{\tau}\right)$ tends to
infinity.

\item As stated in Remark~\ref{rem:ball}, we do not know whether $\left(\Prob_{[0,1]^2},\tau\right)$
admits a uniform embedding into Hilbert space.

\item The present paper rules out the ``low distortion approach" to
nearest neighbor search in the Earthmover metric via embeddings into
$L_1$. However, it might still be possible to find {\em nearest
neighbor preserving embeddings} into $L_1$ in the sense
of~\cite{IN05}.

\item On the more ``applied side", as stated in the introduction,
there is a possibility that the embedding of Theorem~\ref{thm:upper}
behaves better than the theoretical distortion guarantee of $O(\log
n)$ in ``real life" situations, since it is often the case that the
bulk of the Fourier spectrum is concentrated on a sparse set of
frequencies. Additionally, it might be worthwhile to ``thin out"
some frequencies of the given set of images before embedding into
$L_1$ (and then using the known $L_1$ nearest neighbor search
databases). It would be interesting to carry out such ``tweaking" of
our algorithm in a more experimental setting.

\end{enumerate}

\remove{ \noindent{\bf Acknowledgements}  We thank the nice
Continental Airlines representative for changing our seats on a
flight to Texas A\&M University, so that we can work together on
this paper. The second named author was supported in part by the
Israel Science Foundation, and has been emotionally supporting the
first named author for many years (for which the first named author
is eternally grateful).}

\bigskip
\bigskip

\noindent{\bf Acknowledgements.} We are grateful to David Jerison
and Terry Tao for several helpful suggestions. This work was
carried out while the second named author was a long-term visitor
at the Theory Group of Microsoft Research.

\bigskip
\bibliography{emd}

\def\cprime{$'$}
\begin{thebibliography}{10}

\bibitem{AMM85}
I.~Aharoni, B.~Maurey, and B.~S. Mityagin.
\newblock Uniform embeddings of metric spaces and of {B}anach spaces into
  {H}ilbert spaces.
\newblock {\em Israel J. Math.}, 52(3):251--265, 1985.

\bibitem{AFHKTT04}
A.~Archer, J.~Fakcharoenphol, C.~Harrelson, R.~Krauthgamer, K.~Talwar, and
  E.~Tardos.
\newblock Approximate classification via earthmover metrics.
\newblock In {\em SODA '04: Proceedings of the fifteenth annual ACM-SIAM
  symposium on Discrete algorithms}, pages 1079--1087. Society for Industrial
  and Applied Mathematics, 2004.

\bibitem{ALN05}
S.~Arora, J.~R. Lee, and A.~Naor.
\newblock Euclidean distortion and the sparsest cut.
\newblock In {\em STOC '05: Proceedings of the thirty-seventh annual ACM
  symposium on Theory of computing}, pages 553--562, New York, NY, USA, 2005.
  ACM Press.

\bibitem{BL00}
Y.~Benyamini and J.~Lindenstrauss.
\newblock {\em Geometric nonlinear functional analysis. {V}ol. 1}, volume~48 of
  {\em American Mathematical Society Colloquium Publications}.
\newblock American Mathematical Society, Providence, RI, 2000.

\bibitem{BBPW01}
E.~Berkson, J.~Bourgain, A.~Pe{\l}czynski, and M.~Wojciechowski.
\newblock Canonical {S}obolev projections of weak type {$(1,1)$}.
\newblock {\em Mem. Amer. Math. Soc.}, 150(714):viii+75, 2001.

\bibitem{Bourgain87}
J.~Bourgain.
\newblock Remarks on the extension of {L}ipschitz maps defined on discrete sets
  and uniform homeomorphisms.
\newblock In {\em Geometrical aspects of functional analysis (1985/86)}, volume
  1267 of {\em Lecture Notes in Math.}, pages 157--167. Springer, Berlin, 1987.

\bibitem{BS88}
J.~Bourgain and S.~J. Szarek.
\newblock The {B}anach-{M}azur distance to the cube and the
  {D}voretzky-{R}ogers factorization.
\newblock {\em Israel J. Math.}, 62(2):169--180, 1988.

\bibitem{Cha02}
M.~S. Charikar.
\newblock Similarity estimation techniques from rounding algorithms.
\newblock In {\em STOC '02: Proceedings of the thiry-fourth annual ACM
  symposium on Theory of computing}, pages 380--388. ACM Press, 2002.

\bibitem{CKNZ01}
C.~Chekuri, S.~Khanna, J.~Naor, and L.~Zosin.
\newblock Approximation algorithms for the metric labeling problem via a new
  linear programming formulation.
\newblock In {\em SODA '01: Proceedings of the twelfth annual ACM-SIAM
  symposium on Discrete algorithms}, pages 109--118. Society for Industrial and
  Applied Mathematics, 2001.

\bibitem{DIIM04}
M.~Datar, N.~Immorlica, P.~Indyk, and V.~S. Mirrokni.
\newblock Locality-sensitive hashing scheme based on p-stable distributions.
\newblock In {\em SoCG '04: Proceedings of the Twentieth Annual Symposium on
  Computational Geometry}, pages 253--262, New York, NY, USA, 2004. ACM Press.

\bibitem{deLeeuw65}
K.~de~Leeuw.
\newblock On {$L\sb{p}$} multipliers.
\newblock {\em Ann. of Math. (2)}, 81:364--379, 1965.

\bibitem{G-CRdeF85}
J.~Garc{\'{\i}}a-Cuerva and J.~L. Rubio~de Francia.
\newblock {\em Weighted norm inequalities and related topics}, volume 116 of
  {\em North-Holland Mathematics Studies}.
\newblock North-Holland Publishing Co., Amsterdam, 1985.
\newblock Notas de Matem\'atica [Mathematical Notes], 104.

\bibitem{Giann95}
A.~Giannopoulos.
\newblock {A note on the Banach-Mazur distance to the cube.}
\newblock In {\em {Lindenstrauss, J. (ed.) et al., Geometric aspects of
  functional analysis. Israel seminar (GAFA) 1992-94. Basel: Birkhauser. Oper.
  Theory, Adv. Appl. 77, 67-73 }}. 1995.

\bibitem{GRT00}
L.~J. Guibas, Y.~Rubner, and C.~Tomassi.
\newblock The earth mover's distance as a metric for image retrieval.
\newblock {\em International Journal of Computer Vision}, 40(2):99--121, 2000.

\bibitem{GRT98}
L.~J. Guibas, Y.~Rubner, and C.~Tomassi.
\newblock A metric for distributions with applications to image databases.
\newblock In {\em ICCV '98: Proceedings of the Sixth International Conference
  on Computer Vision}, pages 59--66, 2003.

\bibitem{Hein80}
S.~Heinrich.
\newblock {Ultraproducts in Banach space theory.}
\newblock {\em J. Reine Angew. Math.}, 313:72--104, 1980.

\bibitem{HM82}
S.~Heinrich and P.~Mankiewicz.
\newblock Applications of ultrapowers to the uniform and {L}ipschitz
  classification of {B}anach spaces.
\newblock {\em Studia Math.}, 73(3):225--251, 1982.

\bibitem{Hor60}
L.~H{\"o}rmander.
\newblock {Estimates for translation invariant operators in $L^p$ spaces.}
\newblock {\em Acta Math.}, 104:93--140, 1960.

\bibitem{Ind01}
P.~Indyk.
\newblock Algorithmic applications of low-distortion geometric embeddings.
\newblock In {\em 42nd Annual Symposium on Foundations of Computer Science},
  pages 10--33. IEEE Computer Society, 2001.

\bibitem{Ind00}
P.~Indyk.
\newblock Stable distributions, pseudorandom generators, embeddings and data
  stream computation.
\newblock In {\em 41st Annual Symposium on Foundations of Computer Science},
  pages 189--197. IEEE Computer Society, 2001.

\bibitem{Ind04-stoc}
P.~Indyk.
\newblock Algorithms for dynamic geometric problems over data streams.
\newblock In {\em STOC '04: Proceedings of the thirty-sixth annual ACM
  symposium on Theory of computing}, pages 373--380, New York, NY, USA, 2004.
  ACM Press.

\bibitem{Ind04-handbook}
P.~Indyk.
\newblock Nearest neighbors in high-dimensional spaces.
\newblock In {\em Handbook of discrete and computational geometry, second
  edition}, pages 877--892. CRC Press, Inc., Boca Raton, FL, USA, 2004.

\bibitem{IndMat04-handbook}
P.~Indyk and J.~Matou{\v{s}}ek.
\newblock Low distortion embeddings of finite metric spaces.
\newblock In {\em Handbook of discrete and computational geometry, second
  edition}, pages 177--196. CRC Press, Inc., Boca Raton, FL, USA, 2004.

\bibitem{InM98}
P.~Indyk and R.~Motwani.
\newblock Approximate nearest neighbors: towards removing the curse of
  dimensionality.
\newblock In {\em STOC '98: Proceedings of the Thirtieth Annual ACM Symposium
  on Theory of Computing}, pages 604--613, New York, NY, USA, 1998. ACM Press.

\bibitem{IN05}
P.~Indyk and A.~Naor.
\newblock Nearest neighbor preserving embeddings.
\newblock Manuscript. Available at\\ {\footnotesize
  \url{http://research.microsoft.com/research/theory/naor/homepage%20files/low%
dim-journal.pdf}}, 2005.

\bibitem{IT03}
P.~Indyk and N.~Thaper.
\newblock Fast image retrieval via embeddings.
\newblock In {\em ICCV '03: Proceedings of the 3rd International Workshop on
  Statistical and Computational Theories of Vision}, 2003.

\bibitem{Kak41}
S.~Kakutani.
\newblock {Concrete representation of abstract (L)-spaces and the mean ergodic
  theorem.}
\newblock {\em Ann. Math. (2)}, 42:523--537, 1941.

\bibitem{Kalton85}
N.~J. Kalton.
\newblock Banach spaces embedding into {$L\sb 0$}.
\newblock {\em Israel J. Math.}, 52(4):305--319, 1985.

\bibitem{Katz04}
Y.~Katznelson.
\newblock {\em An introduction to harmonic analysis}.
\newblock Cambridge Mathematical Library. Cambridge University Press,
  Cambridge, third edition, 2004.

\bibitem{KN05}
S.~Khot and A.~Naor.
\newblock Nonembeddability theorems via {F}ourier analysis.
\newblock In {\em 46th Annual IEEE Symposium on Foundations of Computer Science
  (FOCS'05)}.
\newblock To appear. Available at\\ {\footnotesize
  \url{http://research.microsoft.com/research/theory/naor/homepage%20files/non%
embed-final-new.pdf}}.

\bibitem{KV05}
S.~Khot and N.~Vishnoi.
\newblock The unique games conjecture, integrality gap for cut problems, and
  embeddability of negative type metrics into ${L}_1$.
\newblock In {\em 46th Annual IEEE Symposium on Foundations of Computer Science
  (FOCS'05)}.
\newblock To appear.

\bibitem{Kis75}
S.~Kislyakov.
\newblock {Sobolev imbedding operators and the nonisomorphism of certain Banach
  spaces.}
\newblock {\em Funct. Anal. Appl.}, 9:290--294, 1975.

\bibitem{LR69}
J.~Lindenstrauss and H.~P. Rosenthal.
\newblock The {${\mathscr L}\sb{p}$} spaces.
\newblock {\em Israel J. Math.}, 7:325--349, 1969.

\bibitem{LTII77}
J.~Lindenstrauss and L.~Tzafriri.
\newblock {\em Classical {B}anach spaces. {II}}, volume~97 of {\em Ergebnisse
  der Mathematik und ihrer Grenzgebiete [Results in Mathematics and Related
  Areas]}.
\newblock Springer-Verlag, Berlin, 1979.
\newblock Function spaces.

\bibitem{jiriproblems}
J.~Matou{\v{s}}ek.
\newblock Open problems on embeddings of finite metric spaces.
\newblock {\em Discrete Comput. Geom.}
\newblock To appear. Available at {\footnotesize
  \url{http://kam.mff.cuni.cz/$\sim$matousek/metrop.ps.gz}}.

\bibitem{Mat01}
J.~Matou{\v{s}}ek.
\newblock {\em Lectures on discrete geometry}, volume 212 of {\em Graduate
  Texts in Mathematics}.
\newblock Springer-Verlag, New York, 2002.

\bibitem{MN05}
M.~Mendel and A.~Naor.
\newblock Metric cotype.
\newblock Preprint, 2005.

\bibitem{Mih56}
S.~G. Mihlin.
\newblock On the multipliers of {F}ourier integrals.
\newblock {\em Dokl. Akad. Nauk SSSR (N.S.)}, 109:701--703, 1956.

\bibitem{Nikisin72}
E.~M. Niki{\v{s}}in.
\newblock A resonance theorem and series in eigenfunctions of the {L}aplace
  operator.
\newblock {\em Izv. Akad. Nauk SSSR Ser. Mat.}, 36:795--813, 1972.

\bibitem{Pel89}
A.~Pelczynski.
\newblock {Boundedness of the canonical projection for Sobolev spaces generated
  by finite families of linear differential operators.}
\newblock In {\em {Analysis at Urbana. Vol. 1: Analysis in function spaces,
  Proc. Spec. Year Mod. Anal./Ill. 1986-87, Lond. Math. Soc. Lect. Note Ser.
  137, 395-415 }}. 1989.

\bibitem{PW03}
A.~Pe{\l}czy{\'n}ski and M.~Wojciechowski.
\newblock Sobolev spaces.
\newblock In {\em Handbook of the geometry of Banach spaces, Vol.\ 2}, pages
  1361--1423. North-Holland, Amsterdam, 2003.

\bibitem{PWR89}
S.~Peleg, M.~Werman, and H.~Rom.
\newblock A unified approach to the change of resolution: space and gray-level.
\newblock {\em IEEE Transactions on Pattern Analysis and Machine Intelligence},
  11(7):739--742, 1989.

\bibitem{Ran04}
N.~L. Randrianarivony.
\newblock Characterization of quasi-{B}anach spaces which coarsely embed into a
  {H}ilbert space.
\newblock Manuscript, 2004.

\bibitem{Ribe76}
M.~Ribe.
\newblock {On uniformly homeomorphic normed spaces.}
\newblock {\em Ark. Mat.}, 14:237--244, 1976.

\bibitem{Rock94}
D.~N. Rockmore.
\newblock Efficient computation of {F}ourier inversion for finite groups.
\newblock {\em J. Assoc. Comput. Mach.}, 41(1):31--66, 1994.

\bibitem{Rudin87}
W.~Rudin.
\newblock {\em {Real and complex analysis. 3rd ed.}}
\newblock {New York, NY: McGraw-Hill.}, 1987.

\bibitem{Stein70}
E.~Stein.
\newblock {\em {Singular integrals and differentiability properties of
  functions}}.
\newblock {Princeton University Press. XIV. Princeton, N.J. }, 1970.

\bibitem{Stein93}
E.~M. Stein.
\newblock {\em Harmonic analysis: real-variable methods, orthogonality, and
  oscillatory integrals}, volume~43 of {\em Princeton Mathematical Series}.
\newblock Princeton University Press, Princeton, NJ, 1993.
\newblock With the assistance of Timothy S. Murphy, Monographs in Harmonic
  Analysis, III.

\bibitem{ST89}
S.~J. Szarek and M.~Talagrand.
\newblock An ``isomorphic'' version of the {S}auer-{S}helah lemma and the
  {B}anach-{M}azur distance to the cube.
\newblock In {\em Geometric aspects of functional analysis (1987--88)}, volume
  1376 of {\em Lecture Notes in Math.}, pages 105--112. Springer, Berlin, 1989.

\bibitem{Tal90}
M.~Talagrand.
\newblock Embedding subspaces of {$L\sb 1$} into {$l\sp N\sb 1$}.
\newblock {\em Proc. Amer. Math. Soc.}, 108(2):363--369, 1990.

\bibitem{Tom89}
N.~Tomczak-Jaegermann.
\newblock {\em Banach-{M}azur distances and finite-dimensional operator
  ideals}, volume~38 of {\em Pitman Monographs and Surveys in Pure and Applied
  Mathematics}.
\newblock Longman Scientific \& Technical, Harlow, 1989.

\bibitem{Tor04}
A.~Torchinsky.
\newblock {\em Real-variable methods in harmonic analysis}.
\newblock Dover Publications Inc., Mineola, NY, 2004.
\newblock Reprint of the 1986 original [Dover, New York; MR0869816].

\bibitem{Villani03}
C.~Villani.
\newblock {\em Topics in optimal transportation}, volume~58 of {\em Graduate
  Studies in Mathematics}.
\newblock American Mathematical Society, Providence, RI, 2003.

\bibitem{Woj96}
P.~Wojtaszczyk.
\newblock {\em Banach spaces for analysts}, volume~25 of {\em Cambridge Studies
  in Advanced Mathematics}.
\newblock Cambridge University Press, Cambridge, 1991.

\bibitem{Woz96}
K.~Wo{\'z}niakowski.
\newblock A new proof of the restriction theorem for weak type {$(1,1)$}
  multipliers on {$\bold R\sp n$}.
\newblock {\em Illinois J. Math.}, 40(3):479--483, 1996.

\bibitem{Zyg02}
A.~Zygmund.
\newblock {\em {Trigonometric series. Volumes I and II combined. With a
  foreword by Robert Fefferman. 3rd ed.}}
\newblock {Cambridge Mathematical Library. Cambridge: Cambridge University
  Press. xiii}, 2002.

\end{thebibliography}
\bibliographystyle{abbrv}

\end{document}